\documentclass[floatfix,showpacs,showkeys,pra,aps,twocolumn]{revtex4}

 \usepackage{bm}
 \usepackage{amsmath}
 \usepackage{amssymb}
 \usepackage{latexsym}
 \usepackage{amsfonts}
 \usepackage{epsfig}
 \usepackage{psfrag}

 \newcommand{\ket}[1]{\left|#1\right>}
 \newcommand{\bra}[1]{\left<#1\right|}
 \newcommand{\expval}[1]{\left< #1 \right>}

 \newcommand{\nn}{\nonumber\\}
 
 \newcommand{\f}[1]{\mbox{\boldmath$#1$}}

 \newcommand{\bea}{\begin{eqnarray}}
 \newcommand{\ea}{\end{eqnarray}}
 \newcommand{\eea}{\end{eqnarray}}
 \newcommand{\ord}{{\cal O}}
 \newcommand{\trace}[1]{{\rm Tr}\left\{ #1 \right\}}
 \newcommand{\traceB}[1]{{\rm Tr_B}\left\{ #1 \right\}}
 \newcommand{\abs}[1]{{\left| #1 \right|}}
 
 \newcommand{\HS}{H_{\rm S}}
 \newcommand{\HB}{H_{\rm B}}
 \newcommand{\HI}{H_{\rm SB}}
 \newcommand{\RS}{\rho_{\rm S}}
 \newcommand{\RB}{\rho_{\rm B}}
 \newcommand{\sinc}[1]{{\rm sinc}\left[#1\right]}

 \newcommand{\RSI}{\f{\RS^0}}
 \newcommand{\HEO}{\f{H_{\rm eff}^{\tau,1}}}
 \newcommand{\HET}{\f{H_{\rm eff}^{\tau,2}}}
 
 \newcommand{\sgn}[1]{{\rm sgn}\left( #1 \right)}
 \newcommand{\GR}{\Gamma_{\rm R}}
 \newcommand{\GL}{\Gamma_{\rm L}}

 \newcommand{\ed}{\varepsilon_{\rm d}}
 \newcommand{\er}{\varepsilon_{\rm R}}
 \newcommand{\el}{\varepsilon_{\rm L}}
 \newcommand{\dr}{\delta_{\rm R}}
 \newcommand{\dl}{\delta_{\rm L}}

\begin{document}

\title{Systematic Perturbation Theory for  Dynamical Coarse-Graining}

\author{Gernot Schaller$^*$, Philipp Zedler, and Tobias Brandes}

\affiliation{Institut f\"ur Theoretische Physik, Technische Universit\"at Berlin, Hardenbergstr. 36, 10623 Berlin, Germany}

\begin{abstract}
We demonstrate how the dynamical coarse-graining approach can be systematically extended to higher orders in the coupling between
system and reservoir.
Up to second order in the coupling constant we explicitly show that dynamical coarse-graining unconditionally preserves
positivity of the density matrix -- even for bath density matrices that are not in equilibrium and
also for time-dependent system Hamiltonians.
By construction, the approach correctly captures the short-time dynamics, i.e., it is suitable to
analyze non-Markovian effects.
We compare the dynamics with the exact solution for highly non-Markovian systems and
find a remarkable quality of the coarse-graining approach.
The extension to higher orders is straightforward but rather tedious.
The approach is especially useful for bath correlation functions of simple structure and
for small system dimensions.
\end{abstract}



\maketitle


\section{Introduction}

The insight that quantum computers may solve certain problems such as number factoring \cite{shor2004a}
and database search \cite{grover1997a} more efficiently than conventional computers has given rise to 
the field of quantum information (for an overview see e.g. \cite{nielsen2000}).
The conventional paradigm of quantum computation relies on unitary operations that act on low-dimensional 
subspaces of the $2^n$-dimensional Hilbertspace of $n$ two-level systems -- conventionally called qubits.
Unfortunately, the dynamics of open quantum systems is not always unitary \cite{breuer2002}, such that 
the impact of decoherence has to be taken into account.
This problem also affects alternative schemes such as one-way
\cite{raussendorf2001a,raussendorf2003a}, holonomic \cite{pachos2001a} or adiabatic \cite{farhi2001} 
quantum computation.
Beyond this, the study of decoherence effects is of general interest in the control of quantum systems.

Often, the dynamics of open quantum systems is analyzed within the Born-Markov approximation scheme
\cite{breuer2002,schlosshauer2007}.
An important criticism raised against this scheme is that it does not generally preserve 
positivity of the reduced density matrix \cite{zhao2002a,stenholm2004a,whitney2008a,taj2008a}, which however
is necessary for its probability interpretation \cite{pechukas1994a}.
In addition, the Born-Markov master equation cannot be expected to yield good results for short times, 
which may in the context of quantum computation for example lead to false error estimates on required 
gate operation times etc \cite{divincenzo2000a,divincenzo2005a}.

A possible resolution for the latter problem is to study non-markovian master equations (that 
explicitly depend on the density matrix at all previous times via a memory kernel).
However, except for some special cases \cite{maniscalco2007a}, non-markovian master equations are also not
guaranteed to preserve positivity, and corresponding counterexamples can be easily constructed \cite{zedler2008a}.
Technically, master equations with memory can for example be solved efficiently when the bath correlation 
functions can be approximated by a few decaying exponentials \cite{kleinekathoefer2004a}.
In the general case, they are however difficult if not impossible to solve analytically.
This difficulty transfers to the numerical solution as well: In order to evolve the density matrix at time $t$, one
would generally have to evaluate the solution at all previous times $t'<t$, which corresponds to significant computational
and storage efforts.

It is therefore interesting to investigate alternatives such as the dynamical coarse-graining (DCG) 
approach.
Recently, it been analyzed up to second order in the
system-reservoir coupling constant (Born approximation) \cite{schaller2008a}.
Instead of solving a single quantum master equation, the coarse-graining approach defines a
continuous set of master equations
\bea
\dot{\f{\RS^\tau}} &=& \f{{\cal L}^\tau} \f{\RS^\tau}(t)
\eea
parametrized by the coarse-graining time $\tau$ and then interpolates through the
set of solutions at $t=\tau$ 
\bea
\f{\bar{\RS}}(t) &=& e^{\f{{\cal L}^t} t} \RSI\,.
\eea
Since the Liouville superoperators $\f{{\cal L}^\tau}$ are of Lindblad form \cite{lindblad1976a}
for all $\tau > 0$, the second-order dynamical coarse-graining 
approach (DCG2) preserves positivity of the density matrix at all times \cite{schaller2008a}.
Note that in the general case, the above solution cannot be obtained by solving a single Lindblad form 
master equation merely equipped with time-dependent coefficients and should therefore be regarded
as truly non-Markovian \cite{wolf2008a}.
The conventional Born-Markov-Secular limit is obtained by the limit $\tau\to\infty$, i.e., 
$\f{\dot{\RS}^{\rm BMS}}= \f{{\cal L}^\infty} \f{\RS^{\rm BMS}}$, whereas in the short-time limit, 
the exact full solution is approximated.
In addition, it was found for some simple examples considered in \cite{schaller2008a} that in the 
weak coupling limit, the method approximated the results of the non-Markovian master equation 
for all times remarkably well.

The purpose of the present paper is two-fold. 
By introducing coarse graining in the interaction picture in section \ref{Sdcgint} we 
rigorously demonstrate that the method will approximate the exact solution for short times 
by construction, i.e., the method is suitable to study non-Markovian effects.
By including higher orders in the coupling constant, the agreement between coarse-graining and
exact solution can be further improved.
In addition, we show that up to second order the method unconditionally preserves positivity of the
density matrix, i.e., even for bath density matrices that do not commute with the bath Hamiltonian
or/and for time-dependent system Hamiltonians.
We will give several examples for finite-size ''baths'' (subsections \ref{SStwospin1} and \ref{SStwospin2}), 
the spin-boson model in subsection \ref{SSspinboson}, and we also consider fermionic models with transport
in subsection \ref{SStransport}.


\section{DCG in the Interaction Picture}\label{Sdcgint}


\subsection{Preliminaries}

We consider systems where the time-independent Hamiltonian can be divided into three parts
\bea
H = \HS + \HI + \HB\,,
\eea
where $\HS$ denotes the system Hamiltonian, $\HB$ the bath (reservoir) Hamiltonian (with 
$\left[\HS, \HB\right]=0$), and
\bea\label{Ehamint}
\HI = \lambda \sum_\alpha A_\alpha \otimes B_\alpha
\eea
couples the two by system ($A_\alpha$) and bath ($B_\alpha$) operators.
Note that thereby one has by construction $\left[A_\alpha, B_\beta\right]=0$ 
(see section \ref{SStransport} for obtaining such a decomposition for fermionic 
systems with transport).

Note that hermiticity of $\HI=\HI^\dagger$ imposes some constraints on the coupling operators.
For example, it is always possible to 
perform a suitable redefinition of operators by splitting into hermitian and anti-hermitian parts 
($A_\alpha=A_\alpha^{\rm H}+A_\alpha^{\rm A}$ and $B_\alpha=B_\alpha^{\rm H}+B_\alpha^{\rm A}$, 
for system and bath operators, respectively) to obtain
$\HI=\frac{1}{2}\left(\HI+\HI^\dagger\right)=\sum_\alpha\left[A_\alpha^{\rm H} B_\alpha^{\rm H}
- i A_\alpha^{\rm A} i B_\alpha^{\rm A}\right]$,
such that one can always assume hermitian coupling operators
$\tilde{A}_\alpha=\tilde{A}_\alpha^\dagger$ as well as $\tilde{B}_\alpha=\tilde{B}_\alpha^\dagger$ \cite{breuer2002}.
For the sake of convenience however, we will not assume this form here unless stated otherwise.
We will use $\lambda < 1$ as a perturbation parameter ($\alpha$-dependent coupling constants can be
absorbed in the operator definitions).

In the interaction picture (where we will denote all operators by bold symbols)
\bea
\f{\rho}(t) &=& e^{+i(\HS+\HB) t} \rho(t) e^{-i (\HS+\HB) t}\,,\nn
\f{A_\alpha}(t) &=& e^{+i\HS t} A_\alpha e^{-i \HS t}\,,\nn
\f{B_\alpha}(t) &=& e^{+i\HB t} B_\alpha e^{-i \HB t}
\eea
the von-Neumann equation reads
\bea
\dot{\f{\rho}} = -i \left[ \f{\HI}(t), \f{\rho}(t)\right]\,,
\eea
which is formally solved by 
\mbox{$\f{\rho}(t) = \f{U}(t) \rho_0 \f{U}^\dagger(t)$}.


\subsection{Perturbative Expansion}

The time evolution operator in the interaction picture is governed by 
$\dot{\f{U}} = -i \f{\HI}(t) \f{U}(t)$, which can be solved iteratively.
We can define the truncated time evolution operator in the interaction picture via
\bea
\f{U_n}(t) &=& \sum_{k=0}^n (-i)^k \int\limits_0^t \f{\HI}(t_1) \ldots \f{\HI}(t_k)\times\nn
&&\times \Theta(t_1 - t_2) \ldots \Theta(t_{k-1}-t_k) dt_1 \ldots dt_k \,,
\eea
where the time-ordering is expressed by Heaviside step functions.
The above operator is unitary up to order of $\lambda^n$ (assuming that $\f{\HI}=\ord\{\lambda\})$, 
i.e., \mbox{$\f{U_n}(t)\f{U_n}^\dagger(t)=\f{1}+\ord\{\lambda^{n+1}\}$}.
Specifically, one has up to fourth order
\bea
\f{U_4}(t) &=& 
\f{1} - i \lambda {\cal V}_1(t) - \lambda^2 {\cal V}_2(t) + i \lambda^3 {\cal V}_3(t) + \lambda^4 {\cal V}_4(t)\,,
\eea
where we can use Eqn. (\ref{Ehamint}) to find for the operators
\bea
{\cal V}_1(t) &\equiv& \sum_\alpha \int\limits_0^t dt_1 \f{A_\alpha}(t_1)  \f{B_\alpha}(t_1)\,,\nn
{\cal V}_2(t) &\equiv& \sum_{\alpha\beta} \int\limits_0^t dt_1 dt_2 \Theta(t_1 - t_2) \f{A_\alpha}(t_1) \f{B_\alpha}(t_1)\times\nn
&&\times \f{A_\beta}(t_2) \f{B_\beta}(t_2)\,,\nn
{\cal V}_3(t) &\equiv& \sum_{\alpha\beta\gamma} \int\limits_0^t dt_1 dt_2 dt_3 \Theta(t_1 - t_2) \Theta(t_2 - t_3) \times\nn
&&\times\f{A_\alpha}(t_1) \f{B_\alpha}(t_1) \f{A_\beta}(t_2) \f{B_\beta}(t_2) \f{A_\gamma}(t_3) \f{B_\gamma}(t_3)\,,\nn
{\cal V}_4(t) &\equiv& \sum_{\alpha\beta\gamma\delta} \int\limits_0^t \Theta(t_1 - t_2)\Theta(t_2 - t_3)\Theta(t_3 - t_4)\times\nn
&&\times\f{A_\alpha}(t_1) \f{B_\alpha}(t_1) \f{A_\beta}(t_2) \f{B_\beta}(t_2)\times\nn
&&\times \f{A_\gamma}(t_3) \f{B_\gamma}(t_3) \f{A_\delta}(t_4) \f{B_\delta}(t_4)dt_1 dt_2 dt_3 dt_4 \,.
\eea

Using these expressions in the formal solution of the density matrix and collecting all terms of the same order we obtain
\bea
\f{\rho}(t) &=& \f{\rho}_0 - i \lambda\left[- \f{\rho}_0 {\cal V}_1^\dagger(t) + {\cal V}_1(t) \f{\rho}_0 \right]\nn 
&&+ \lambda^2\left[- \f{\rho}_0 {\cal V}_2^\dagger(t) + {\cal V}_1(t) \f{\rho}_0 {\cal V}_1^\dagger(t) - {\cal V}_2(t) \f{\rho}_0\right]\nn
&&-i \lambda^3 \Big[ \f{\rho}_0 {\cal V}_3^\dagger(t) - {\cal V}_1(t) \f{\rho}_0 {\cal V}_2^\dagger(t)\nn
&&+ {\cal V}_2(t) \f{\rho}_0 {\cal V}_1^\dagger(t) - {\cal V}_3(t) \f{\rho}_0\Big]\nn
&&+ \lambda^4 \Big[\f{\rho}_0 {\cal V}_4^\dagger(t) - {\cal V}_1(t) \f{\rho}_0 {\cal V}_3^\dagger(t) + {\cal V}_2(t) \f{\rho}_0 {\cal V}_2^\dagger(t)\nn
&&- {\cal V}_3(t) \f{\rho}_0 {\cal V}_1^\dagger(t) + {\cal V}_4(t) \f{\rho}_0\Big] + \ord\{\lambda^5\}\,.
\eea
In order to obtain the reduced density matrix, we have to perform the trace over the bath degrees of freedom. 
We assume that at $t_0=0$ the density matrix factorizes such that we have
$\RSI=\traceB{\f{\rho_0}}$.
Then we can define $\f{\RS}(t)=\traceB{\f{\rho}(t)}$ and calculate the reduced density matrix at time $t$
\bea\label{Epertex}
\f{\RS}(t) &=& \RSI - i \lambda \traceB{- \RSI \RB^0 {\cal V}_1^\dagger + {\cal V}_1 \RSI \RB^0}\nn
&&+ \lambda^2 \traceB{ - \RSI \RB^0 {\cal V}_2^\dagger + {\cal V}_1 \RSI \RB^0 {\cal V}_1^\dagger - {\cal V}_2 \RSI \RB^0}\nn
&&-i \lambda^3 {\rm Tr_B}\Big\{ \RSI \RB^0 {\cal V}_3^\dagger - {\cal V}_1 \RSI \RB^0 {\cal V}_2^\dagger\nn
&& + {\cal V}_2 \RSI \RB^0 {\cal V}_1^\dagger - {\cal V}_3 \RSI \RB^0\Big\}\nn
&&+ \lambda^4 {\rm Tr_B}\Big\{\RSI \RB^0 {\cal V}_4^\dagger - {\cal V}_1 \RSI \RB^0 {\cal V}_3^\dagger 
+ {\cal V}_2 \RSI \RB^0 {\cal V}_2^\dagger\nn
&&- {\cal V}_3 \RSI \RB^0 {\cal V}_1^\dagger + {\cal V}_4 \RSI \RB^0\Big\} + \ord\{\lambda^5\}\nn
&\equiv& \RSI + \lambda \f{{\cal T}_1^t}\RSI + \lambda^2 \f{{\cal T}_2^t}\RSI + \lambda^3 \f{{\cal T}_3^t}\RSI\nn
&& + \lambda^4 \f{{\cal T}_4^t}\RSI + \ord\{\lambda^5\}\,,
\eea
where we can evaluate the right-hand side by using the reservoir correlation functions, see below.


\subsection{Bath Correlation Functions}

Since we do neither assume {\em a priori} that the coupling operators are hermitian
nor that $\left[\HB, \RB^0\right]=0$, it is necessary to generalize the correlation functions.
Denoting the index of a hermitian conjugate coupling operator with an overbar, we define for the first and second order
\bea\label{Ecorrfunc_nonherm}
C_\alpha(t_1) &\equiv& \traceB{B_\alpha(t_1) \RB^0}\,,\nn
C_{\bar\alpha}(t_1) &\equiv& \traceB{B_\alpha^\dagger(t_1) \RB^0}\,,\nn
C_{\alpha\beta}(t_1, t_2) &\equiv& \traceB{B_\alpha(t_1) B_\beta(t_2) \RB^0}\,,\nn
C_{\alpha\bar\beta}(t_1, t_2) &\equiv& \traceB{B_\alpha(t_1) B_\beta^\dagger(t_2) \RB^0}\,,
\eea
and similarly for higher-orders. 
In terms of these quantities, the r.h.s. of Eqn. (\ref{Epertex}) can be easily evaluated.
\begin{widetext}
Specifically, we obtain
\bea\label{Epertall}
\f{{\cal T}_1^t}\RSI &=& -i \sum_\alpha \int\limits_0^t dt_1 \left[- C_{\bar\alpha}(t_1) \RSI \f{A_\alpha^\dagger}(t_1) + C_\alpha(t_1) \f{A_\alpha}(t_1) \RSI \right]\,,\nn
\f{{\cal T}_2^t}\RSI &=& \sum_{\alpha\beta} \int\limits_0^t dt_1 dt_2 
\Big[ -C_{\bar\alpha\bar\beta}(t_1, t_2) \Theta(t_2-t_1) \RSI \f{A_\alpha^\dagger}(t_1) \f{A_\beta^\dagger}(t_2)\nn
&&+C_{\bar\alpha\beta}(t_1, t_2) \f{A_\beta}(t_2) \RSI \f{A_\alpha^\dagger}(t_1)
-C_{\alpha\beta}(t_1, t_2) \Theta(t_1-t_2)  \f{A_\alpha}(t_1) \f{A_\beta}(t_2) \RSI
\Big]\,,\nn
\f{{\cal T}_3^t}\RSI &=& -i \sum_{\alpha\beta\gamma} \int\limits_0^t dt_1 dt_2 dt_3 
\Big[+ C_{\bar\alpha\bar\beta\bar\gamma}(t_1, t_2, t_3) \Theta(t_3-t_2) \Theta(t_2-t_1) \RSI \f{A_\alpha^\dagger}(t_1) \f{A_\beta^\dagger}(t_2) \f{A_\gamma^\dagger}(t_3)\nn
&&- C_{\bar\alpha\bar\beta\gamma}(t_1, t_2, t_3) \Theta(t_2-t_1) \f{A_\gamma}(t_3) \RSI \f{A_\alpha^\dagger}(t_1) \f{A_\beta^\dagger}(t_2)
+C_{\bar\alpha\beta\gamma}(t_1, t_2, t_3) \Theta(t_2-t_3) \f{A_\beta}(t_2) \f{A_\gamma}(t_3) \RSI \f{A_\alpha^\dagger}(t_1)\nn
&&-C_{\alpha\beta\gamma}(t_1, t_2, t_3) \Theta(t_1-t_2) \Theta(t_2-t_3) \f{A_\alpha}(t_1) \f{A_\beta}(t_2) \f{A_\gamma}(t_3) \RSI\Big]\,,\nn
\f{{\cal T}_4^t}\RSI &=& \sum_{\alpha\beta\gamma\delta} \int\limits_0^t dt_1 dt_2 dt_3 dt_4
\Big[
+C_{\bar\alpha\bar\beta\bar\gamma\bar\delta}(t_1, t_2, t_3, t_4) \Theta(t_4-t_3)\Theta(t_3-t_2)\Theta(t_2-t_1) 
\RSI \f{A_\alpha^\dagger}(t_1) \f{A_\beta^\dagger}(t_2) \f{A_\gamma^\dagger}(t_3) \f{A_\delta^\dagger}(t_4)\nn
&&
-C_{\bar\alpha\bar\beta\bar\gamma\delta}(t_1, t_2, t_3, t_4) \Theta(t_3-t_2)\Theta(t_2-t_1)
\f{A_\delta}(t_4) \RSI \f{A_\alpha^\dagger}(t_1) \f{A_\beta^\dagger}(t_2) \f{A_\gamma^\dagger}(t_3)\nn
&&
+C_{\bar\alpha\bar\beta\gamma\delta}(t_1, t_2, t_3, t_4) \Theta(t_3-t_4)\Theta(t_2-t_1) 
\f{A_\gamma}(t_3) \f{A_\delta}(t_4) \RSI \f{A_\alpha^\dagger}(t_1) \f{A_\beta^\dagger}(t_2)\nn
&&
-C_{\bar\alpha\beta\gamma\delta}(t_1, t_2, t_3, t_4) \Theta(t_2-t_3)\Theta(t_3-t_4) 
\f{A_\beta}(t_2) \f{A_\gamma}(t_3) \f{A_\delta}(t_4) \RSI \f{A_\alpha^\dagger}(t_1)\nn
&&
+C_{\alpha\beta\gamma\delta}(t_1, t_2, t_3, t_4) \Theta(t_1-t_2)\Theta(t_2-t_3)\Theta(t_3-t_4)
\f{A_\alpha}(t_1) \f{A_\beta}(t_2) \f{A_\gamma}(t_3) \f{A_\delta}(t_4) \RSI \Big]\,.
\eea
\end{widetext}


\subsection{Defining the DCG Liouvillian}\label{SSdcgliouville}

It is evident that one can carry on with the expansion of the time-evolution operator to arbitrary 
order in the coupling constant $\lambda$.
This will evidently yield good results for small $\lambda$ and small times, whereas we would like to
have a master equation valid for small $\lambda$ and also large times.
In the original approach \cite{schaller2008a} it was shown as a supportive fact that for $t=\tau$ the 
DCG2 solution and the approximation (\ref{Epertex})
were equivalent up to $\ord\{\lambda^2\}$.
Here we will demand equivalence between the $n$-th order coarse graining solution and Eqn. (\ref{Epertex}) at $t=\tau$ to define our 
perturbation theory.
Expanding the Liouvillian superoperator as 
\mbox{$\f{{\cal L}^\tau}=\lambda \f{{\cal L}_1^\tau} + \lambda^2 \f{{\cal L}_2^\tau} + \lambda^3 \f{{\cal L}_3^\tau} + \lambda^4 \f{{\cal L}_4^\tau} + \ord\{\lambda^5\}$},
we obtain for the solution of
$\dot{\f{\RS^\tau}}(t) = \f{{\cal L}^\tau} \f{\RS^\tau}(t)$
at time $t=\tau$
\bea\label{Esoldcg}
\f{\RS^\tau}(\tau) &=& 
\Big\{
\f{1}
+\lambda \tau \f{{\cal L}_1^\tau}
+\lambda^2\left[\tau \f{{\cal L}_2^\tau} + \frac{\tau^2}{2} \f{{\cal L}_1^\tau} \f{{\cal L}_1^\tau}\right]\nn
&+&\lambda^3\left[\tau \f{{\cal L}_3^\tau} + \frac{\tau^2}{2} \left(\f{{\cal L}_1^\tau} \f{{\cal L}_2^\tau} + \f{{\cal L}_2^\tau} \f{{\cal L}_1^\tau}\right) + \frac{\tau^3}{6} \f{{\cal L}_1^\tau} \f{{\cal L}_1^\tau} \f{{\cal L}_1^\tau}\right]\nn
&+&\lambda^4\Big[\tau \f{{\cal L}_4^\tau} + \frac{\tau^2}{2} \left(\f{{\cal L}_1^\tau} \f{{\cal L}_3^\tau} + \f{{\cal L}_2^\tau} \f{{\cal L}_2^\tau} + \f{{\cal L}_3^\tau} \f{{\cal L}_1^\tau}\right)\nn 
&+& \frac{\tau^3}{6} \left( \f{{\cal L}_1^\tau} \f{{\cal L}_1^\tau} \f{{\cal L}_2^\tau} + \f{{\cal L}_1^\tau} \f{{\cal L}_2^\tau} \f{{\cal L}_1^\tau} + \f{{\cal L}_2^\tau} \f{{\cal L}_1^\tau} \f{{\cal L}_1^\tau}\right)\nn
&+&\frac{\tau^4}{24}  \f{{\cal L}_1^\tau} \f{{\cal L}_1^\tau} \f{{\cal L}_1^\tau} \f{{\cal L}_1^\tau}\Big]
\Big\}\RSI+\ord\{\lambda^5\}
\eea
We can clearly match this with equation (\ref{Epertex}) evaluated at $t=\tau$ order by order to solve for
\bea\label{Edefliouville}
\f{{\cal L}_1^\tau} \RSI &=& \frac{1}{\tau} \f{{\cal T}_1^\tau} \RSI \,,\nn
\f{{\cal L}_2^\tau} \RSI &=& \frac{1}{\tau} \left\{ \f{{\cal T}_2^\tau} - \left[ \frac{\tau^2}{2} \f{{\cal L}_1^\tau} \f{{\cal L}_1^\tau} \right] \right\}\RSI\,,\nn
\f{{\cal L}_3^\tau} \RSI &=& \frac{1}{\tau} \Big\{ \f{{\cal T}_3^\tau}\nn
&&- \left[ \frac{\tau^2}{2} \left(\f{{\cal L}_1^\tau} \f{{\cal L}_2^\tau} + \f{{\cal L}_2^\tau} \f{{\cal L}_1^\tau}\right) + \frac{\tau^3}{6} \f{{\cal L}_1^\tau} \f{{\cal L}_1^\tau} \f{{\cal L}_1^\tau}\right] \Big\}\RSI\,,\nn
\f{{\cal L}_4^\tau} \RSI &=& \frac{1}{\tau} \Big\{ \f{{\cal T}_4^\tau} 
- \Big[ \frac{\tau^2}{2} \left(\f{{\cal L}_1^\tau} \f{{\cal L}_3^\tau} + \f{{\cal L}_2^\tau} \f{{\cal L}_2^\tau} + \f{{\cal L}_3^\tau} \f{{\cal L}_1^\tau}\right)\nn
&&+ \frac{\tau^3}{6} \left(\f{{\cal L}_1^\tau} \f{{\cal L}_1^\tau} \f{{\cal L}_2^\tau} + \f{{\cal L}_1^\tau} \f{{\cal L}_2^\tau} \f{{\cal L}_1^\tau} + \f{{\cal L}_2^\tau} \f{{\cal L}_1^\tau} \f{{\cal L}_1^\tau}\right)\nn
&&+ \frac{\tau^4}{24} \f{{\cal L}_1^\tau} \f{{\cal L}_1^\tau} \f{{\cal L}_1^\tau} \f{{\cal L}_1^\tau}
\Big] \Big\}\RSI\,,
\eea
where $\f{{\cal T}_1^\tau}, \f{{\cal T}_2^\tau}, \f{{\cal T}_3^\tau}, \f{{\cal T}_4^\tau}$ can be extracted from Eqns. (\ref{Epertall}).
Since these equations have to hold for all initial conditions $\RSI$ we can infer the matrix elements of each Liouvillian by comparing
coefficients of the matrix elements of $\f{\RSI}$.

Equations (\ref{Edefliouville}) define in combination with (\ref{Epertall}) our coarse-graining Liouvillian.
Evidently, we automatically approximate the short-time dynamics of the true solution very well by construction with
this scheme.


\subsection{Unconditional Positivity of DCG2}\label{SSpositivity}

Here we will show that DCG2 always preserves positivity -- regardless whether the first order correlation functions vanish or not.
We do not even require $\left[\HB, \RB^0\right]=0$.
For simplicity we assume hermitian coupling operators $A_\alpha=A_\alpha^\dagger$ and $B_\alpha=B_\alpha^\dagger$.
Then, we obtain from Eqns. (\ref{Epertall})
\bea
\f{{\cal T}_1^\tau}\RS &=& -i \sum_\alpha \int\limits_0^\tau dt_1 C_\alpha(t_1) 
\left[ \f{A_\alpha}(t_1) \RS - \RS \f{A_\alpha}(t_1)\right]\,,\nn
\f{{\cal T}_2^\tau}\RS &=& \sum_{\alpha\beta} \int\limits_0^\tau dt_1 dt_2 C_{\alpha\beta}(t_1, t_2)
\Big[\f{A_\beta}(t_2) \RS \f{A_\alpha}(t_1)\nn
&& - \frac{1}{2} \RS \f{A_\alpha}(t_1) \f{A_\beta}(t_2)- \frac{1}{2} \f{A_\alpha}(t_1) \f{A_\beta}(t_2) \RS\Big]\nn
&&-i \sum_{\alpha\beta} \frac{1}{2i} \int\limits_0^\tau C_{\alpha\beta}(t_1, t_2) \sgn{t_1-t_2}\times\nn
&&\times\left[ \f{A_\alpha}(t_1) \f{A_\beta}(t_2) , \RS\right] dt_1 dt_2\,,
\eea
where we have used $\Theta(x)=\frac{1}{2}\left[1+\sgn{x}\right]$ and $\sgn{-x}=-\sgn{x}$ in the last line.
From the first of the above equations we obtain that the first order Liouvillian just generates a unitary evolution
\bea\label{Eliouville_first}
\f{{\cal L}_1^\tau}\RS &=& -i \left[ \frac{1}{\tau} \sum_\alpha \int\limits_0^\tau C_\alpha(t_1) \f{A_\alpha}(t_1) dt_1, \RS \right]\nn
&\equiv& -i \left[\HEO, \RS\right]\,,
\eea
where hermiticity of the Lamb-shift Hamiltonian follows directly from hermiticity of the coupling operators (which also implies real-valued first
order correlation functions).
In addition, we obtain from consecutive application
\bea
\frac{1}{2} \f{{\cal T}_1^\tau}\left[\f{{\cal T}_1^\tau}\RS\right] &=& \sum_{\alpha\beta} \int\limits_0^\tau dt_1 dt_2 C_\alpha(t_1) C_\beta(t_2)\times\nn
&&\times\Big[\f{A_\beta}(t_2) \RS \f{A_\alpha}(t_1)\nn
&&- \frac{1}{2}\left\{ \RS, \f{A_\alpha}(t_1) \f{A_\beta}(t_2)\right\}\Big]\,.
\eea
This defines the second order Liouvillian as
\bea\label{Eliouville_second}
\f{{\cal L}_2^\tau}\RS &=& -i \Big[ \frac{1}{2\tau i} \sum_{\alpha\beta} \int\limits_0^\tau C_{\alpha\beta}(t_1, t_2) \sgn{t_1-t_2}\times\nn
&&\times \f{A_\alpha}(t_1) \f{A_\beta}(t_2) dt_1 dt_2, \RS\Big]\nn
&&+\frac{1}{\tau} \sum_{\alpha\beta} \int\limits_0^\tau dt_1 dt_2 \left[C_{\alpha\beta}(t_1, t_2)-C_\alpha(t_1)C_\beta(t_2)\right]
\times\nn
&&\left[\f{A_\beta}(t_2) \RS \f{A_\alpha}(t_1) - \frac{1}{2} \left\{\f{A_\alpha}(t_1) \f{A_\beta}(t_2),  \RS\right\}\right]\,.
\eea
The first commutator term induces a unitary evolution where hermiticity of the corresponding effective Hamiltonian 
follows directly from $C_{\beta\alpha}^*(t_2, t_1) = C_{\alpha\beta}(t_1, t_2)$.
However, in contrast to the standard Born-Markov-secular approximation \cite{breuer2002} here we have in general
\mbox{$\left[\HET, \HS\right] \neq 0$}.
In order to see that the second line corresponds to a Lindblad dissipator, we insert identities at suitable
places $\f{1}=\sum_a \ket{a}\bra{a}$ to obtain
\begin{widetext}
\bea
\f{{\cal L}_2^\tau}\RS &=& -i \left[ \HET, \RS\right]
+ \sum_{ab,cd} \gamma_{ab,cd}^{\tau,2} 
\left[L_{ab} \RS L_{cd}^\dagger - \frac{1}{2} \left\{L_{cd}^\dagger L_{ab}, \RS\right\}\right]\,,\nn
\gamma_{ab,cd}^{\tau,2} &=& 
\frac{1}{\tau} \sum_{\alpha\beta} \int\limits_0^\tau \left[C_{\alpha\beta}(t_1, t_2) - C_\alpha(t_1) C_\beta(t_2)\right]
\bra{a} \f{A_\beta}(t_2) \ket{b} \bra{c} \f{A_\alpha}(t_1) \ket{d}^* dt_1 dt_2\,,
\eea
where we have abbreviated the operators $L_{ab}=\ket{a}\bra{b}$.
The dampening matrix elements can be most conveniently evaluated in the energy eigenbasis $\HS \ket{a} = E_a \ket{a}$.

However, independent from the basis choice it remains to be shown that the dampening matrix 
is positive semidefinite to get a Lindblad form.
In order to see this, we calculate with Eqn. (\ref{Ecorrfunc_nonherm})
\bea
\sum_{abcd} x_{ab}^* \gamma_{ab,cd}^{\tau,2} x_{cd} &=& \frac{1}{\tau} \sum_{abcd} \sum_{\alpha\beta} x_{ab}^* x_{cd} \int\limits_0^\tau dt_1 dt_2 C_{\alpha\beta}(t_1, t_2) 
\bra{a} \f{A_\beta}(t_2) \ket{b} \bra{c} \f{A_\alpha}(t_1) \ket{d}^*\nn
&&-\frac{1}{\tau} \sum_{abcd} \sum_{\alpha\beta} x_{ab}^* x_{cd} \int\limits_0^\tau dt_1 dt_2 C_\alpha(t_1) C_\beta(t_2) 
\bra{a} \f{A_\beta}(t_2) \ket{b} \bra{c} \f{A_\alpha}(t_1) \ket{d}^*\nn
&=& \frac{1}{\tau} \traceB{\left[\sum_{cd\alpha} \int\limits_0^\tau \f{B_\alpha}(t_1) x_{cd} \bra{c} \f{A_\alpha}(t_1) \ket{d}^* dt_1 \right]
\left[\sum_{ab\beta} \int\limits_0^\tau \f{B_\beta}(t_2) x_{ab}^* \bra{a} \f{A_\beta}(t_2) \ket{b} dt_2 \right] \RB^0}\nn
&&-\frac{1}{\tau} \traceB{\sum_{cd\alpha} \int\limits_0^\tau \f{B_\alpha}(t_1) x_{cd} \bra{c} \f{A_\alpha}(t_1) \ket{d}^* dt_1\RB^0}
\traceB{\sum_{ab\beta} \int\limits_0^\tau \f{B_\beta}(t_2) x_{ab}^* \bra{a} \f{A_\beta}(t_2) \ket{b} dt_2 \RB^0}\nn
&\equiv& \frac{1}{\tau}\left[ \traceB{K^\dagger(\tau) K(\tau) \RB^0} - \traceB{K^\dagger(\tau) \RB^0} \traceB{K(\tau) \RB^0}\right]\nn
&=& \frac{1}{\tau}\left[ \traceB{K^\dagger(\tau) K(\tau) \RB^0} - \abs{\traceB{K(\tau) \RB^0}}^2\right]\,.
\eea 
\end{widetext}
Whereas the first term in the last line appears positive, one might fear that positivity can be spoiled by the existence
of the additional second term.
However, we can bound the second term via the Cauchy-Schwarz trace inequality \cite{liu1995a}
$\abs{\trace{AB}}^2 \le \trace{A^\dagger A} \trace{B^\dagger B}$ with 
$A=K(\tau)\sqrt{\RB^0}$ and $B=\sqrt{\RB^0}$ (which exists as $\RB^0$ is positive semidefinite)
\bea
\abs{\traceB{K(\tau) \RB^0}}^2 &=& \abs{\traceB{K(\tau) \sqrt{\RB^0} \sqrt{\RB^0}}}^2\nn
&\le& \traceB{\sqrt{\RB^0} K^\dagger(\tau) K(\tau) \sqrt{\RB^0}}\times\nn
&&\times \traceB{\sqrt{\RB^0}\sqrt{\RB^0}}\nn
&=& \traceB{K^\dagger(\tau) K(\tau) \RB^0}\,.
\eea
Remembering that $\traceB{K^\dagger(\tau) K(\tau) \RB^0} \ge 0$ for any operator $K(\tau)$ we therefore 
obtain for $\tau \ge 0$
\bea
\sum_{abcd} x_{ab}^* \gamma_{ab,cd}^{\tau,2} x_{cd} \ge 0\,,
\eea
i.e., we have generated a Lindblad form master equation.
This result goes beyond ref \cite{schaller2008a} in several aspects:
Not only is the case $C_\alpha(t_1)\neq 0$ considered but in addition, 
we do not constrain ourselves to bath density matrices in thermal equilibrium, i.e., one also has positivity for 
$\left[\RB^0, \HB \right] \neq 0$\,.
It is an interesting avenue of further research to compare DCG with other methods within
the context of nonequilibrium environments \cite{emary2008a}.
Beyond that, all of the above arguments go through if the system Hamiltonian is time-dependent.
In this case, the coupling operators in the interaction picture have to obey
$\dot{\f{A_\alpha}}=+i \left[\HS(t), \f{A_\alpha}(t)\right]$, such that the challenge is then
to calculate the matrix elements $\bra{a} \f{A_\alpha}(t)\ket{b}$.

Under the assumptions $C_\alpha(t)=0$ (no first order correlation functions), 
$C_{\alpha\beta}(t_1, t_2) = C_{\alpha\beta}(t_1-t_2) \equiv \traceB{\f{B_\alpha}(t_1-t_2) B_\beta \RB^0}$ 
(reservoir in thermal equilibrium) we can insert the Fourier transforms of 
$C_{\alpha\beta}(t_1-t_2)$ and $C_{\alpha\beta}(t_1-t_2) \sgn{t_1-t_2}$.
If in addition the system Hamiltonian is time-independent, we may calculate the time integrals
analytically.
Then, we can make use of the identity for discrete $a$, $b$ (see e.g. appendix F of ref. \cite{schaller2008a})
\bea\label{Edeltafunc}
\lim_{\tau \to \infty} \tau \sinc{\frac{(\omega + a) \tau}{2}} \sinc{\frac{(\omega + b)\tau}{2}} \sim 2 \pi \delta_{ab} \delta(\omega + a)
\eea
to calculate the large time limit of the DCG2 approach.
In complete analogy to ref \cite{schaller2008a} we obtain the Born-Markov secular approximation \cite{breuer2002} 
in this limit.

Unfortunately, the unconditional preservation of positivity is not preserved by higher orders within DCG (although of
course, in the weak coupling limit the nice properties of DCG2 will dominate).


\section{Examples}

In the following, we will test the DCG approach with simple examples for which at least in 
special cases an analytical solution exists.
For finite-size reservoirs the correlation functions are non-decaying and these systems
are inherently non-Markovian (exhibiting for example recurrences), cf. the examples in 
subsections \ref{SStwospin1} and \ref{SStwospin2}.
For quasi-continuous reservoirs we will compare the performance of the DCG approach with
the Born-Markov approximation, see subsections \ref{SSspinboson} and \ref{SStransport}.


\subsection{DCG2 for two spins}\label{SStwospin1}

We consider a highly non-markovian system (S) by using a very small reservoir (B), namely just a single further spin
\bea
\HS &=& \omega \sigma^z_{\rm S}\,,\qquad \HB = \Omega \sigma^z_{\rm B}\,,\\
\HI &=& \lambda \vec{\sigma}_{\rm S} \cdot \vec{\sigma}_{\rm B} 
= \lambda \left[ \sigma^x_{\rm S} \otimes \sigma^x_{\rm B} 
+ \sigma^y_{\rm S} \otimes \sigma^y_{\rm B} 
+ \sigma^z_{\rm S} \otimes \sigma^z_{\rm B} \right]\,,\nonumber
\eea
i.e., the index of the coupling operators runs from one to three.
Note that all coupling operators are hermitian, such that we may omit overbars and daggers in Eqn. (\ref{Epertall}).
We assume that the initial bath density matrix is diagonal in order to simplify all expressions
$\RB^0 = \left(
\begin{array}{cc}
\RB^{00} & 0\nn
0 & 1-\RB^{00}
\end{array}
\right)$.
The exact solution can be obtained by exponentiating the Hamiltonian and tracing out the bath spin (not shown).
As in subsection (\ref{SSpositivity}) we decompose the Liouville operator into unitary and non-unitary contributions,
where we have first and second order contributions in the unitary action of decoherence $\f{H_{\rm eff}^\tau}=\f{H_{\rm eff}^{\tau,1}}+\f{H_{\rm eff}^{\tau,2}}$ 
and second order contributions for the dissipative action $\gamma_{ab, cd}^\tau=\gamma_{ab, cd}^{\tau,2}$.

Transforming the coupling operators into the interaction picture we obtain
$\f{B_1}(t) = \cos(2\Omega t) \sigma^x_{\rm B} - \sin(2\Omega t) \sigma^y_{\rm B}$, 
$\f{A_1}(t) = \cos(2\omega t) \sigma^x_{\rm S} - \sin(2\omega t) \sigma^y_{\rm S}$,
$\f{B_2}(t) = \cos(2\Omega t) \sigma^y_{\rm B} + \sin(2\Omega t) \sigma^x_{\rm B}$, 
$\f{A_2}(t) = \cos(2\omega t) \sigma^y_{\rm S} + \sin(2\omega t) \sigma^x_{\rm S}$,
$\f{B_3}(t) = \sigma^z_{\rm B}$, and
$\f{A_3}(t) = \sigma^z_{\rm S}$.
From this, we obtain the time-independent first order correlation functions
\bea
C_1(t) = 0\,,\qquad C_2(t) = 0\,,\qquad C_3(t) = 2 \RB^{00}-1\,,
\eea
which yield for the first order Lamb shift Hamiltonian from Eqn. (\ref{Eliouville_first})
\bea
\HEO &=& \lambda (2 \RB^{00}-1) \sigma^z_{\rm S}\,.
\eea

The non-vanishing second order correlation functions equate to
\bea
C_{11} &=& \cos[2 (t_1 - t_2) \Omega] - i (1 - 2 \RB^{00}) \sin[2 (t_1 - t_2) \Omega]\,,\nn
C_{12} &=& -i (1 - 2 \RB^{00}) \cos[2 (t_1 - t_2) \Omega] - \sin[2 (t_1 - t_2) \Omega]\,,\nn
C_{21} &=& i (1 - 2 \RB^{00}) \cos[2 (t_1 - t_2) \Omega] + \sin[2 (t_1 - t_2) \Omega]\,,\nn
C_{22} &=& \cos[2 (t_1 - t_2) \Omega] - i (1 - 2 \RB^{00}) \sin[2 (t_1 - t_2) \Omega]\,,\nn
C_{33} &=& 1\,,
\eea
where we have omitted the time-dependencies for brevity.
This can be inserted in the expression for the second order Lamb-shift Hamiltonian
in Eqn. (\ref{Eliouville_second}) to yield
\bea
\HET &=& \frac{2\lambda^2}{\Omega-\omega} \left\{ 1 - \sinc{2\tau \left(\Omega - \omega\right)}\right\}\times\nn
&&\times\left[\left(\RB^{00}-\frac{1}{2}\right)\f{1}_{\rm S}-\frac{1}{2}\sigma^z_{\rm S}\right]\,,
\eea
which commutes with the system Hamiltonian.

\begin{figure}[ht]
\begin{tabular}{c}
\includegraphics[height=6cm,clip=true]{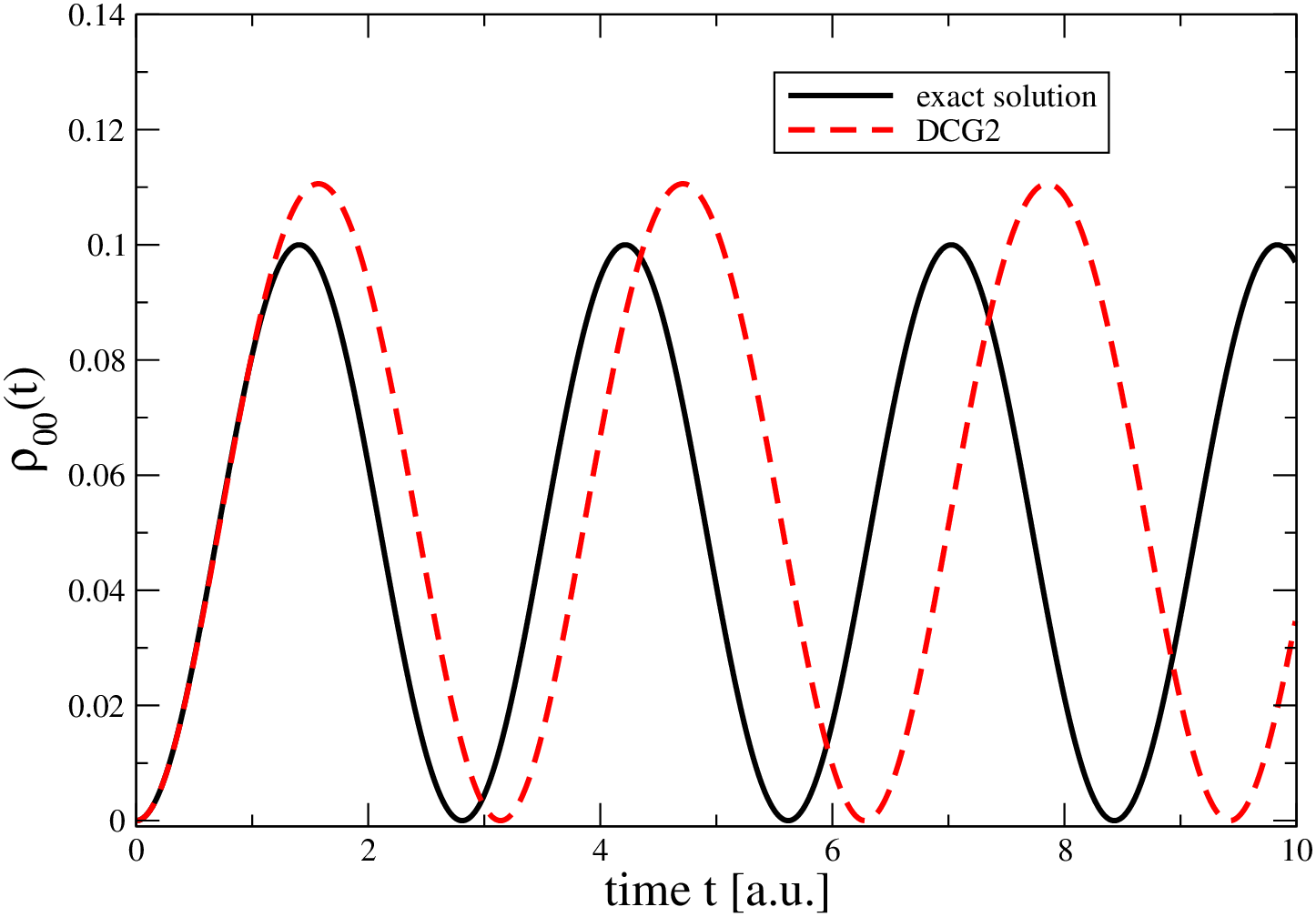}\\
\includegraphics[height=6cm,clip=true]{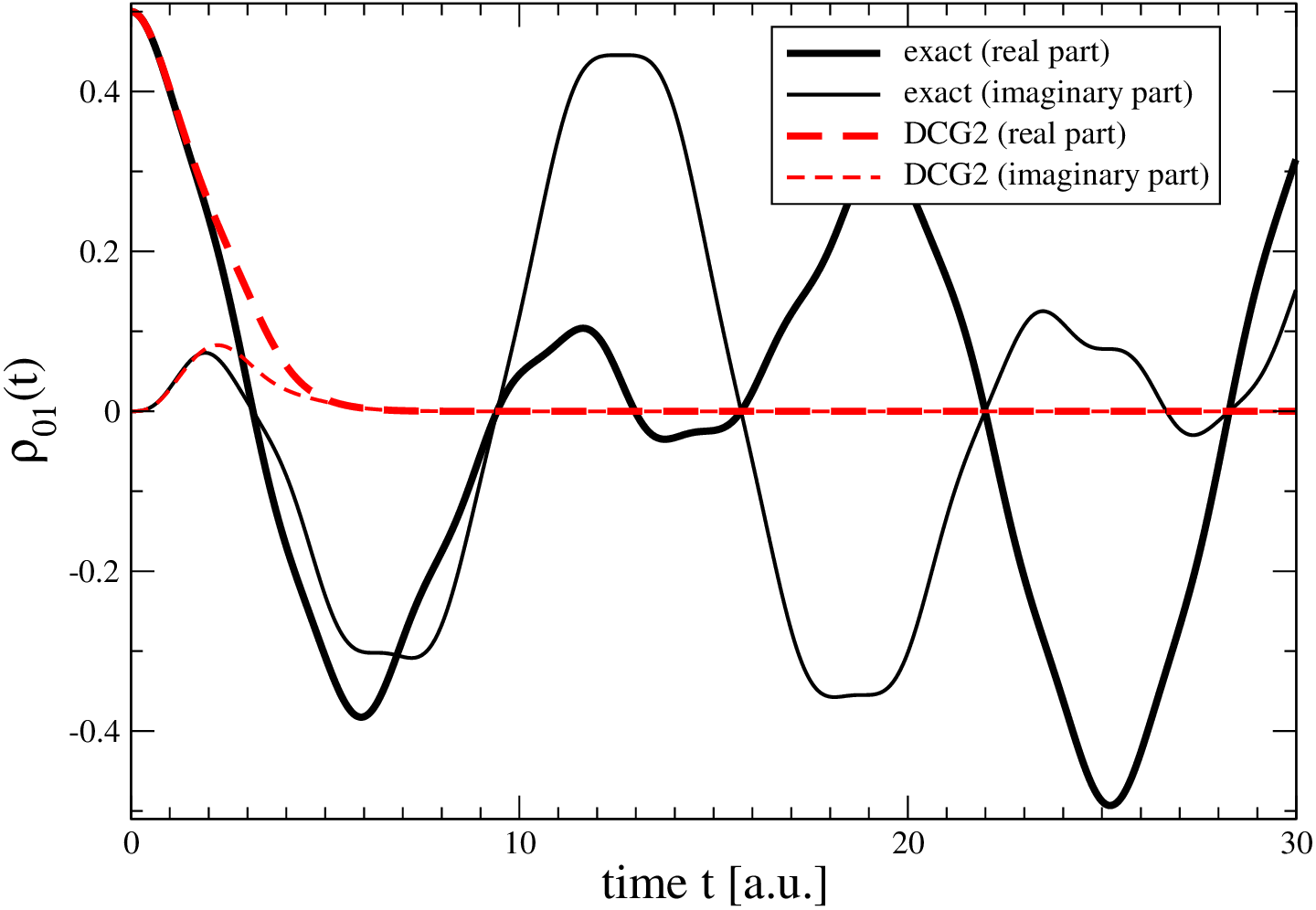}
\end{tabular}
\caption{\label{Fspindcg2}[Color Online]
Comparison of exact (solid black) and DCG2 (dashed red) solutions for the diagonal (a) and off-diagonal (b) 
matrix elements of the density matrix.
Imaginary parts are displayed with thin lines.
Naturally, the exact solution displays complete recurrences.
For the diagonal matrix elements, this feature is well reproduced by the DCG2 approach,
whereas for the off-diagonals only the short-time dynamics is well approximated.
The other parameters have been chosen as follows $\lambda=0.25$, $\omega=1.0$, $\Omega=2.0$, and 
$\RB^{00}=0.5$.
}
\end{figure}

The second order dissipative terms must be calculated from the dissipative parts of Eqn. (\ref{Eliouville_second}),
where we obtain for the non-vanishing matrix elements of the dampening matrix
\mbox{$\gamma_{00,00}^{\tau,2} = 4 \lambda^2 \tau (1 - \RB^{00}) \RB^{00}$},
\mbox{$\gamma_{00,11}^{\tau,2} = - 4 \lambda^2 \tau (1 - \RB^{00}) \RB^{00}$},
\mbox{$\gamma_{11,00}^{\tau,2} = - 4 \lambda^2 \tau (1 - \RB^{00}) \RB^{00}$},
\mbox{$\gamma_{11,11}^{\tau,2} = 4 \lambda^2 \tau (1 - \RB^{00}) \RB^{00}$},
\mbox{$\gamma_{01,01}^{\tau,2} = 4 \lambda^2 \tau \RB^{00} {\rm sinc}^2 \left[\tau (\Omega - \omega)\right]$}, and
\mbox{$\gamma_{10,10}^{\tau,2} = 4 \lambda^2 \tau \left(1-\RB^{00}\right) {\rm sinc}^2 \left[\tau (\Omega - \omega)\right]$},
which shows (e.g., by the Gershgorin circle theorem \cite{varga2004}) that $\gamma_{ab,cd}$ is positive semidefinite.
The solution of the coarse-graining master equation 
$\dot{\f{\RS^\tau}}(t) = \f{{\cal L}^\tau} \f{\RS^\tau}(t)$
can be conveniently obtained by exploiting that diagonal and off-diagonal matrix elements decouple.
From the diagonal equations
\bea
\dot{\f{\RS^{00}}}(t) &=& \gamma_{01,01}^{\tau,2} \f{\RS^{11}}(t) - \gamma_{10,10}^{\tau,2} \f{\RS^{00}}(t)\nn
&=& \gamma_{01,01}^{\tau,2} - \left[ \gamma_{01,01}^{\tau,2} + \gamma_{10,10}^{\tau,2} \right] \f{\RS^{00}}(t)\nn
&=& 4 \lambda^2 \tau \RB^{00} {\rm sinc}^2 \left[\tau (\Omega - \omega)\right]\nn
&&- 4 \lambda^2 \tau {\rm sinc}^2 \left[\tau (\Omega - \omega)\right] \f{\RS^{00}}(t)
\eea
we obtain the solution at $\tau=t$ 
\bea
\bra{0} \f{\RS^t}(t) \ket{0} &=& 
\exp\left\{- 4 \frac{\lambda^2}{(\Omega-\omega)^2} \sin^2\left[t (\Omega - \omega)\right]\right\} \RS^{00}(0)\nn
&&+ \left[ 1 - \exp\left\{- \frac{4 \lambda^2 \sin^2\left[t (\Omega - \omega)\right]}{(\Omega-\omega)^2} \right\}\right] \RB^{00}
\,,\nonumber
\eea
which does admit for complete recurrences of the populations, see figure \ref{Fspindcg2} (a).

For the off-diagonal equation
\bea
\dot{\f{\RS^{01}}}(t) &=& \Big[ \gamma_{00,11}^{\tau,2} 
- \frac{1}{2} \left( \gamma_{00,00}^{\tau,2} + \gamma_{01,01}^{\tau,2} + \gamma_{10,10}^{\tau,2} + \gamma_{11,11}^{\tau,2}\right)\nn
&& +i \Big( \bra{1}\HEO \ket{1} + \bra{1} \HET \ket{1} - \bra{0} \HEO \ket{0}\nn
&& - \bra{0} \HET \ket{0} \Big) \Big] \f{\RS^{01}}(t)\nn
&=& \Big\{ - 8 \lambda^2 \tau \RB^{00}(1-\RB^{00}) - 2 \lambda^2 \tau {\rm sinc}^2\left[ \tau (\Omega - \omega)\right]\nn
&&+ i \Big(\frac{2\lambda^2\left(1 - \sinc{2\tau(\Omega - \omega)}\right)}{\Omega - \omega}\nn
&& + 2\lambda (1- 2 \RB^{00})\Big)\Big\} \f{\RS^{01}}(t)
\eea
we obtain the solution
\bea
\bra{0} \f{\RS^t}(t) \ket{1} &=& e^{
- 8 \lambda^2 t^2 \RB^{00}(1-\RB^{00}) - 2 \lambda^2 t^2 {\rm sinc}^2\left[ \tau (\Omega - \omega)\right]}\times\nn
&&e^{+ i t \left[\frac{2\lambda^2\left(1 - \sinc{2t (\Omega - \omega)}\right)}{\Omega - \omega} 
+ 2\lambda (1- 2 \RB^{00})\right]} \f{\RS^{01}}(0)\,,
\eea
where we will generally observe a decay whenever $\RB^{00}(1-\RB^{00}) \neq 0$, see figure \ref{Fspindcg2} (b).


\subsection{DCG4 for two spins}\label{SStwospin2}

In order to keep the calculations for DCG4 very simple, we consider
\bea
\HS = \omega \sigma^z_{\rm S}\,, \qquad \HB = \Omega \sigma^z_{\rm B}\,,\qquad \HI =
 \lambda \sigma^x_{\rm S} \otimes \sigma^z_{\rm B}\,,
\eea
where also here the coupling operators are hermitian.
In this case, the bath correlation functions are all time-independent, which enables a 
convenient calculation of the Liouvillian matrix elements.
The example is of course a bit trivial, since the exact solution for the reduced density matrix
does not depend on $\Omega$.
Note however, that unlike the pure dephasing limit considered in \cite{krovi2007a} this case
still holds some time-dependence that can be found in the system operator in
the interaction picture.

\begin{figure}[ht]
\begin{tabular}{c}
\includegraphics[height=6cm,clip=true]{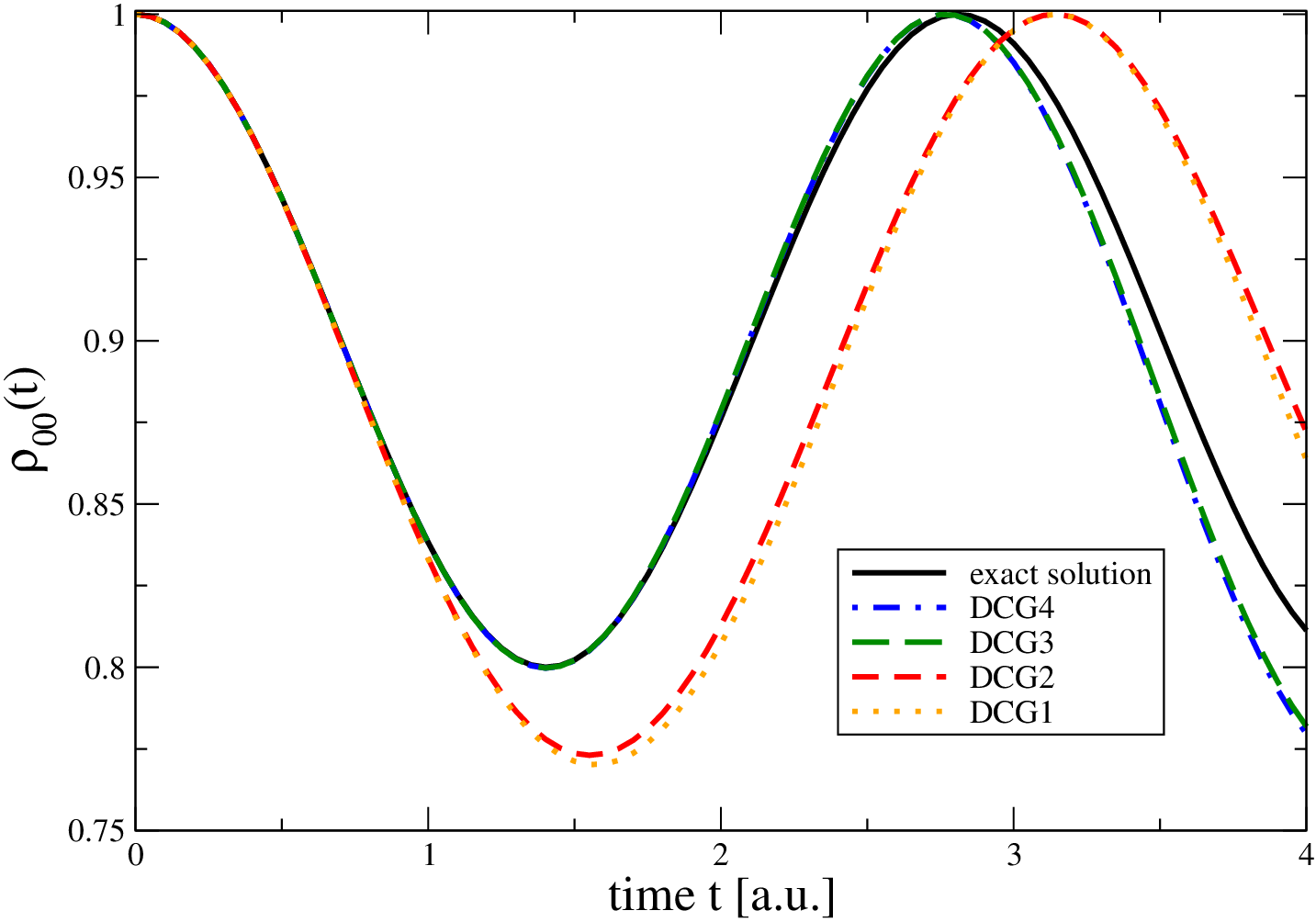}\\
\includegraphics[height=6cm,clip=true]{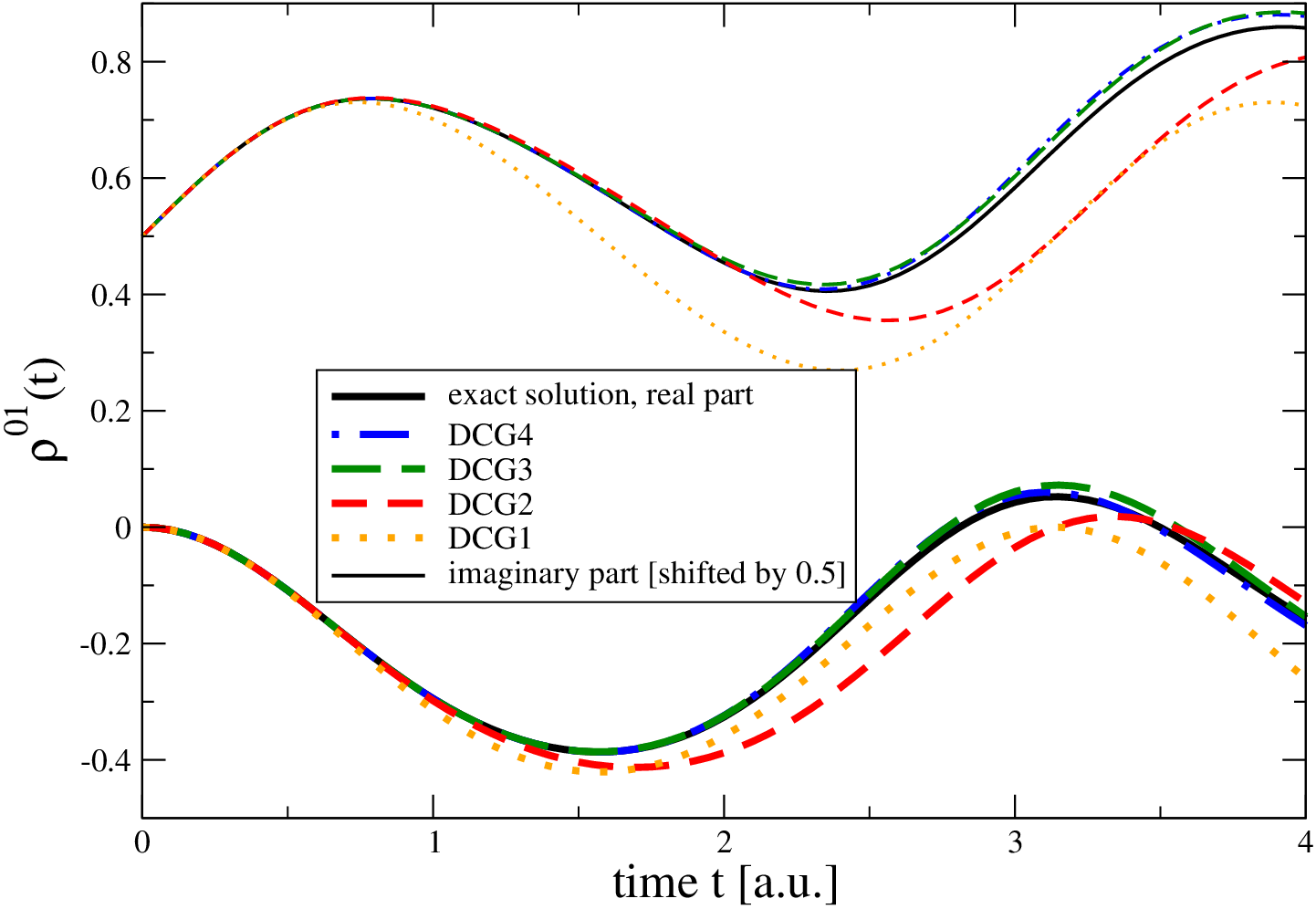}
\end{tabular}
\caption{\label{Fspindcg4}[Color Online]
Comparison of exact (solid black) and DCG1 (dotted orange), DCG2 (dashed red), DCG3 (long-dashed green), and 
DCG4 solutions (dot-dashed lines) for the diagonal (a) and 
off-diagonal (b) matrix elements of the density matrix.
For small times, DCG4 is superior to the coarse-graining methods of smaller accuracy.
By construction, the exact solution shows complete recurrences.
For larger times, all coarse graining methods also display recurrences but all of them
miss the exact solution (not shown).
Parameters have been chosen as $\lambda=0.5$, $\omega=1$, and $\RB^{00}=1$.
}
\end{figure}

The exact solution can be calculated by exponentiating the complete Hamiltonian and tracing out the 
second spin in the solution for the density matrix as in subsection \ref{SStwospin1} and in a similar manner
we determine the DCG1, DCG2, DCG3, and DCG4 solutions by directly determining the $4\times 4$ Liouvillian 
matrix as described in subsection \ref{SSdcgliouville} (not shown).
The solution is then obtained by exponentiating the Liouvillian.
The resulting solution for the diagonals is displayed in figure \ref{Fspindcg4} (a) and 
for the off-diagonals in figure \ref{Fspindcg4} (b).


\subsection{Spin-Boson model}\label{SSspinboson}

We consider a single system spin coupled to a bath of bosonic modes ($\omega_k > 0$)
\bea\label{Ehamspinboson}
\HS &=& \frac{\ed}{2}\left(\f{1} - \sigma^z\right)\,,\qquad
\HB = \sum_k \omega_k \left( b_k^\dagger b_k + \frac{1}{2}\right)\,,\nn
\HI &=& \lambda A \otimes \left[ \sum_k h_k b_k + h_k^* b_k^\dagger\right]\,,
\eea
where for simplicity we have restricted ourselves to the case of single-operator coupling, 
and the operator $A=A^\dagger$ will be specified later-on.

We will consider a thermalized initial bath density matrix
\bea\label{Edmatbath}
\RB^0 = \frac{e^{-\beta \HB}}{\traceB{e^{-\beta \HB}}}
\eea
where $\beta=(k_{\rm B} T)^{-1}$ denotes the inverse reservoir temperature.


\subsubsection{Bath Correlation Functions}

We evaluate the traces in the correlation functions in Eqn. (\ref{Ecorrfunc_nonherm}) in Fock-space,
where the bath density matrix (\ref{Edmatbath}) is diagonal.
By doing so it becomes obvious that the number of creation and annihilation operators in each term of the bath-bath
correlation functions must be balanced for all modes to obtain a nonvanishing result.
Therefore, we conclude (since only one operator is involved, we may omit the indices)
$C(t_1) = 0 = C(t_1,t_2,t_3)$.
In the interaction picture, the annihilation and creation operators transform according to
$\f{b_k}(t) = e^{+i\HB t} b_k e^{-i\HB t} = e^{-i \omega_k t} b_k$ and the hermitian conjugate, respectively.

The second order correlation function then evaluates to
\bea\label{Ecfunc2sp}
C(t_1,t_2) &=& 
\frac{1}{2\pi} \int\limits_{0}^{\infty} d\omega G(\omega) \Big\{ n(\omega) e^{+i \omega (t_1 - t_2)}\nn
&&+ \left[1+ n(\omega)\right] e^{-i \omega (t_1 - t_2)}\Big\}\nn
&=& \frac{1}{2\pi} \int\limits_{-\infty}^{+\infty} \frac{G(\abs{\omega})}{\abs{e^{\beta \omega} - 1}} e^{+i \omega (t_1 - t_2)} d\omega\,,
\eea
where the bosonic occupation number is given by \mbox{$n(\omega) = \frac{1}{e^{\beta \omega} - 1}$}.
In the above equation, we have assumed a quasi-continuous spectral density 
\mbox{$G(\omega)= 2\pi \sum_k \abs{h_k}^2 \delta(\omega - \omega_k)$} to
convert the sum into an integral.
When we parametrize the spectral density as
\bea\label{Edof_spinboson}
G(\omega) = G_0 \omega^S e^{-\omega/\omega_{\rm c}}\,,
\eea
where $\omega_{\rm c}$ denotes a cutoff frequency and the parameter $S$ governs the slope at $\omega=0$, we can obtain
an analytic solution for the correlation function \cite{weiss1993,brandes2005a}
\bea
C(t_1,t_2)
&=& \frac{G_0 \Gamma(1+S)}{2\pi \beta^{1+S}}
\Big[\zeta\left(1+S, \frac{1}{\beta \omega_{\rm c}} + i \frac{(t_1-t_2)}{\beta}\right)\nn
&&+ \zeta\left(1+S, 1+ \frac{1}{\beta \omega_{\rm c}} - i \frac{(t_1-t_2)}{\beta}\right)\Big]
\eea
in terms of generalized Riemann Zeta functions $\zeta(x,y)$.

The next non-vanishing correlation function is fourth order, where we obtain
\bea\label{Ecfunc4sp}
C(t_1, t_2, t_3, t_4) &=& C(t_2,t_3) C(t_1,t_4) + C(t_1,t_3) C(t_2,t_4)\nn
&& + C(t_1,t_2) C(t_3,t_4)\,,
\eea
which can for example be obtained using Wicks theorem (for a special case see also Eqn. (61) in \cite{divincenzo2005a}).


\subsubsection{Pure Dephasing}\label{SSpuredephasing}

The case of pure dephasing $A=\sigma^z$ is exactly solvable \cite{unruh1995a,lidar2001a} and it is known that
DCG2 already yields the exact result \cite{schaller2008a}.
The exact solution predicts time-independent diagonal matrix elements and a decay of the off-diagonal
matrix element according to (in the interaction picture, cf. Eqn. (82) in \cite{lidar2001a}
 in the limit of a continuous bath spectrum)
\bea\label{Edecay_exact}
\f{\rho_{01}}(t) &=& e^{-\Gamma(t)} \f{\rho_{01}}(0)\,,\nn
\Gamma(t) &=& \frac{8\lambda^2}{2\pi} \int\limits_0^\infty G(\omega) \frac{\sin^2(\omega t/2)}{\omega^2} {\coth}\left[\frac{\beta\omega}{2}\right] d\omega\,,
\eea
where the additional factor of $\frac{1}{2\pi}$ in comparison to \cite{schaller2008a} results from a different 
definition of the spectral density $G(\omega)$.
By taking the time derivative of the exact solution density matrix we obtain a closed master equation that is
not of Lindblad form (not even with time-dependent coefficients) but nevertheless must -- as it is exact -- preserve
positivity.
Hence, one would regard this case as truly non-Markovian \cite{wolf2008a}.
The corresponding steady state is also derived by the equation of motion method in appendix \ref{Aspinboson}.
For pure dephasing we have $\f{A}(t)=A=\sigma^z$. 
We observe a decoupled evolution of diagonal and off-diagonal matrix elements of the density matrix.

For the second order contribution we have (using $\Theta(t_1-t_2)+\Theta(t_2-t_1) = 1$) from
Eqn. (\ref{Epertall})
\bea
\f{{\cal T}_2^\tau}\RSI = \int\limits_0^\tau  C(t_1, t_2) \left[ \sigma^z \RSI \sigma^z - \RSI\right] dt_1 dt_2\,,
\eea
from which we obtain
\bea
\bra{0} \f{{\cal T}_2^\tau}\RSI \ket{0} &=& \bra{1} \f{{\cal T}_2^\tau}\RSI \ket{1} = 0\,,\nn
\bra{0} \f{{\cal T}_2^\tau}\RSI \ket{1} &=& -2 \int\limits_0^\tau C(t_1, t_2) dt_1 dt_2 \f{\RS^{01}}\nn
&=&-\frac{8}{2\pi} \int\limits_0^\infty G(\omega) \frac{\sin^2(\omega\tau/2)}{\omega^2} \coth\left[\frac{\beta\omega}{2}\right] d\omega \f{\RS^{01}}\,,\nonumber
\eea
which leads to the same exponential decay as with the exact solution (\ref{Edecay_exact}), i.e., 
as noted earlier \cite{schaller2008a}, DCG2 yields the exact solution in this case.
Note that this example demonstrates explicitly that the DCG2 solution cannot be generally written as
the solution of a single Lindblad-type master equation with time-dependent coefficients, such that it should
be classified as non-Markovian \cite{wolf2008a}.

Using the relation
\bea\label{Eheavi00}
0 &=&
+\Theta(t_4-t_3)\Theta(t_3-t_2)\Theta(t_2-t_1)\nn
&&+\Theta(t_3-t_4)\Theta(t_2-t_1)\nn
&&+\Theta(t_1-t_2)\Theta(t_2-t_3)\Theta(t_3-t_4)\nn
&&-\Theta(t_2-t_3)\Theta(t_3-t_4)
-\Theta(t_3-t_2)\Theta(t_2-t_1)\qquad
\eea
we obtain for the fourth order contribution
\bea
\bra{0} \f{{\cal T}_4^\tau}\RSI \ket{0} &=& \bra{1} \f{{\cal T}_4^\tau}\RSI \ket{1} = 0\,,\nn
\bra{0} \f{{\cal T}_4^\tau}\RSI \ket{1} &=& 2 \f{\RS^{01}} \int\limits_0^\tau dt_1 dt_2 dt_3 dt_4 C(t_1, t_2, t_3, t_4)\times\nn
&&\Big[ \Theta(t_3-t_2)\Theta(t_2-t_1)\nn
&&+ \Theta(t_2-t_3)\Theta(t_3-t_4)\Big]\nn
&=& 2 \int\limits_0^\tau dt_1 dt_2 dt_3 dt_4 C(t_1,t_2) C(t_3,t_4) \f{\RS^{01}}\,,
\eea
where we have exploited the symmetries of the fourth order correlation functions under exchange of the arguments and 
the relation
\bea
2&=& \Big[+\Theta(t_2 - t_3)\Theta(t_3 - t_4)+\Theta(t_3 - t_2)\Theta(t_2 - t_1)\qquad\nn
&&+\Theta(t_1 - t_2)\Theta(t_2 - t_4)+\Theta(t_2 - t_1)\Theta(t_1 - t_3)\nn
&&+\Theta(t_3 - t_2)\Theta(t_2 - t_4)+\Theta(t_2 - t_3)\Theta(t_3 - t_1)\nn
&&+\Theta(t_4 - t_1)\Theta(t_1 - t_2)+\Theta(t_1 - t_4)\Theta(t_4 - t_3)\nn
&&+\Theta(t_3 - t_4)\Theta(t_4 - t_2)+\Theta(t_4 - t_3)\Theta(t_3 - t_1)\nn
&&+\Theta(t_1 - t_4)\Theta(t_4 - t_2)+\Theta(t_4 - t_1)\Theta(t_1 - t_3)\Big]\,.
\eea
The non-vanishing off-diagonal contribution has to be compared with the counter-term arising from the second order
\bea
\frac{1}{2}\bra{0} \f{{\cal T}_2^\tau}[\f{{\cal T}_2^\tau}\RSI] \ket{1} &=& \frac{1}{2} 
\left[-2 \int\limits_0^\tau C(t_1,t_2) dt_1 dt_2\right]^2 \f{\RS^{01}}\nn
&=& 2 \left[\int\limits_0^\tau C(t_1,t_2) dt_1 dt_2\right]\times\nn
&&\left[\int\limits_0^\tau C(t_3,t_4) dt_3 dt_4\right] \f{\RS^{01}}
\eea
Since the diagonal elements of the density matrix are neither affected by 
$\f{{\cal T}_2^\tau}$ nor by $\f{{\cal T}_4^\tau}$ we conclude that we have for pure dephasing
\bea
\f{{\cal T}_4^\tau} = \frac{1}{2} \f{{\cal T}_2^\tau} \f{{\cal T}_2^\tau}\,,
\eea
such that DCG4 will yield the same result as DCG2.
Since DCG2 is already exact for this case, this cancellation
is a strong indicator for the correctness of our fourth order correlation function (\ref{Ecfunc4sp}).


\subsubsection{Dissipation}\label{SSdissipation}

We are now in a position to apply DCG to more interesting coupling operators 
(picking up a time-dependence in the interaction picture) that also affect the
evolution of the diagonal elements of the density matrix.
Transforming $A=\sigma^x$ into the interaction picture we obtain
$\f{A}(t)=\cos(\ed t) \sigma^x + \sin(\ed t) \sigma^y$.
Inserting this result we find for the second order term in Eqn. (\ref{Epertall})
\begin{widetext}
\bea
\bra{0} \f{{\cal T}_2^\tau}[\f{\RS}] \ket{0} &=& \int\limits_0^\tau dt_1 dt_2 C(t_1, t_2)
\Big[-e^{-i\ed(t_1-t_2)} \f{\RS^{00}} + e^{+i\ed(t_1-t_2)} \f{\RS^{11}}\Big]\,,\nn
\bra{0} \f{{\cal T}_2^\tau}[\f{\RS}] \ket{1} &=& \int\limits_0^\tau dt_1 dt_2 C(t_1, t_2)
\Big[-\Theta(t_2-t_1) e^{+i\ed(t_1-t_2)} \f{\RS^{01}}
-\Theta(t_1-t_2) e^{-i\ed(t_1-t_2)} \f{\RS^{01}}
+ e^{-i\ed(t_1+t_2)} \f{\RS^{10}}\Big]
\eea
and similarly for the other terms, such that by arranging the matrix elements of the $2\times 2$ density 
matrix in a 4-dimensional vector as $(\f{\RS^{00}}, \f{\RS^{01}}, \f{\RS^{10}}, \f{\RS^{11}})$
we find the matrix elements of the corresponding $4\times 4$ superoperator to be
\bea
\f{{\cal T}_2^\tau} = \int\limits_0^\tau C(t_1, t_2) 
\left(
\begin{array}{cccc}
-e^{-i\ed(t_1-t_2)} & 0 & 0 & +e^{+i\ed(t_1-t_2)}\\
0 & -e^{-i\ed \abs{t_1-t_2}} & +e^{-i\ed(t_1+t_2)} & 0\\
0 & +e^{+i\ed(t_1+t_2)} & -e^{+i\ed \abs{t_1-t_2}} & 0\\
+e^{-i\ed(t_1-t_2)} & 0 & 0 & -e^{+i\ed(t_1-t_2)}
\end{array}
\right)
dt_1 dt_2\,,
\eea
\end{widetext}
such that we observe a decoupled evolution of diagonal and off-diagonal matrix elements.
Defining
\bea\label{Ebandfilter_sb}
m_{11}(\tau) &\equiv& \left(\f{{\cal T}_2^\tau}\right)_{11}\nn
&=&-\frac{\tau^2}{2\pi} \int\limits_{-\infty}^{+\infty} \frac{G(\abs{\omega})}{\abs{e^{\beta\omega}-1}}  
{\rm sinc}^2\left[(\omega-\ed)\frac{\tau}{2}\right] d\omega\,,\nn
m_{14}(\tau) &\equiv& \left(\f{{\cal T}_2^\tau}\right)_{14}\nn
&=&+\frac{\tau^2}{2\pi} \int\limits_{-\infty}^{+\infty} \frac{G(\abs{\omega})}{\abs{e^{\beta\omega}-1}}
{\rm sinc}^2\left[(\omega+\ed)\frac{\tau}{2}\right] d\omega\,,\nn
m_{41}(\tau) &\equiv& \left(\f{{\cal T}_2^\tau}\right)_{41} = -m_{11}(\tau)\,,\nn
m_{44}(\tau) &\equiv& \left(\f{{\cal T}_2^\tau}\right)_{44} = -m_{14}(\tau)
\eea
we obtain the second order solution for the diagonals (using trace conservation)
\bea
\rho_{00}^\tau(\tau) &=& \rho_{00}^0 \exp\left[\lambda^2\left(m_{11}(\tau)-m_{14}(\tau)\right)\right]\nn
&&+\frac{1-\exp\left[\lambda^2\left(m_{11}(\tau)-m_{14}(\tau)\right)\right]}{1-\frac{m_{11}(\tau)}{m_{14}(\tau)}}\,.
\eea
In Eqn. (\ref{Ebandfilter_sb}) it is already obvious that for finite times $\tau$ all frequencies will contribute to
the matrix elements $m_{ij}(\tau)$ in contrast to the Markov approximation, where only $G(\ed)$ is relevant.
With using that the bandfilter sinc functions transform into Dirac-Delta functions in Eqn. (\ref{Edeltafunc}) 
we can perform the limit $\tau\to\infty$ to obtain the steady state
\bea\label{Ess_sb}
\rho_{00}^\infty = \frac{1}{1+e^{-\beta \ed}}\,,
\eea
which corresponds to the thermalized system density matrix that consistent with our expectations, 
compare also appendix \ref{Aspinboson}.
The same stationary state can also be obtained by using the method of equation of motion and
truncating correlations at second order between system and reservoir.

The Markovian limit is obtained by using
\bea\label{Emarkovlimit}
m_{11}^{\rm MK}(t) = t \lim_{\tau\to\infty} \frac{1}{\tau} m_{11}(\tau) = -t G(\abs{\ed}) \abs{n(+\ed)}\,,\nn
m_{14}^{\rm MK}(t) = t \lim_{\tau\to\infty} \frac{1}{\tau} m_{14}(\tau) = +t G(\abs{\ed}) \abs{n(-\ed)}\,,
\eea
where we have again used identity (\ref{Edeltafunc}) in Eqn. (\ref{Ebandfilter_sb}).
Evidently, the above equation leads to the same steady state as the Born-Markov approximation (\ref{Ess_sb}).

By virtue of Eqn. (\ref{Eheavi00}) and using 
that $\left[C(t_1,t_2,t_3,t_4)=C(t_4,t_3,t_2,t_1)\right]^*$ we obtain for the diagonal part
\bea
\bra{0} \f{{\cal T}_4^\tau}\RSI\ket{0} &=& 
2\int\limits_0^\tau \Re\left\{ C(t_1,t_2,t_3,t_4) e^{-i \ed(t_1-t_2+t_3-t_4)}\right\}\times\nn
&&\Theta(t_2-t_3)\Theta(t_3-t_4) dt_1 dt_2 dt_3 dt_4 \f{\RS^{00}}\nn
&&- 2\int\limits_0^\tau \Re\left\{ C(t_1,t_2,t_3,t_4) e^{+i \ed(t_1-t_2+t_3-t_4)} \right\}\times\nn
&&\Theta(t_2-t_3)\Theta(t_3-t_4) dt_1 dt_2 dt_3 dt_4 \f{\RS^{11}}\nn
&\equiv& p_{11}(\tau) \f{\RS^{00}} + p_{14}(\tau) \f{\RS^{11}}\,.
\eea
This result has to be combined with the counterterm arising from the squared second order contribution.
Defining
\bea
\tilde p_{11}(\tau) &=& \lambda^2 m_{11}(\tau)-\frac{\lambda^4}{2}\left[m_{11}(\tau)m_{11}(\tau) + m_{14}(\tau)m_{41}(\tau)\right]\nn
&&+\lambda^4 p_{11}(\tau)\,,\nn
\tilde p_{14}(\tau) &=& \lambda^2 m_{14}(\tau)-\frac{\lambda^4}{2}\left[m_{11}(\tau)m_{14}(\tau) + m_{14}(\tau)m_{44}(\tau)\right]\nn
&&+\lambda^4 p_{14}(\tau)\,,
\eea
we therefore obtain the fourth order solution
\bea
\rho_{00}^\tau(\tau) &=& \rho_{00}^0 \exp\left[\tilde p_{11}(\tau)- \tilde p_{14}(\tau)\right]\nn
&&+\frac{1-\exp\left[\tilde p_{11}(\tau)- \tilde p_{14}(\tau)\right]}{1-\frac{\tilde p_{11}(\tau)}{\tilde p_{14}(\tau)}}\,.
\eea

\begin{figure}[ht]
\begin{tabular}{c}
\includegraphics[height=6cm,clip=true]{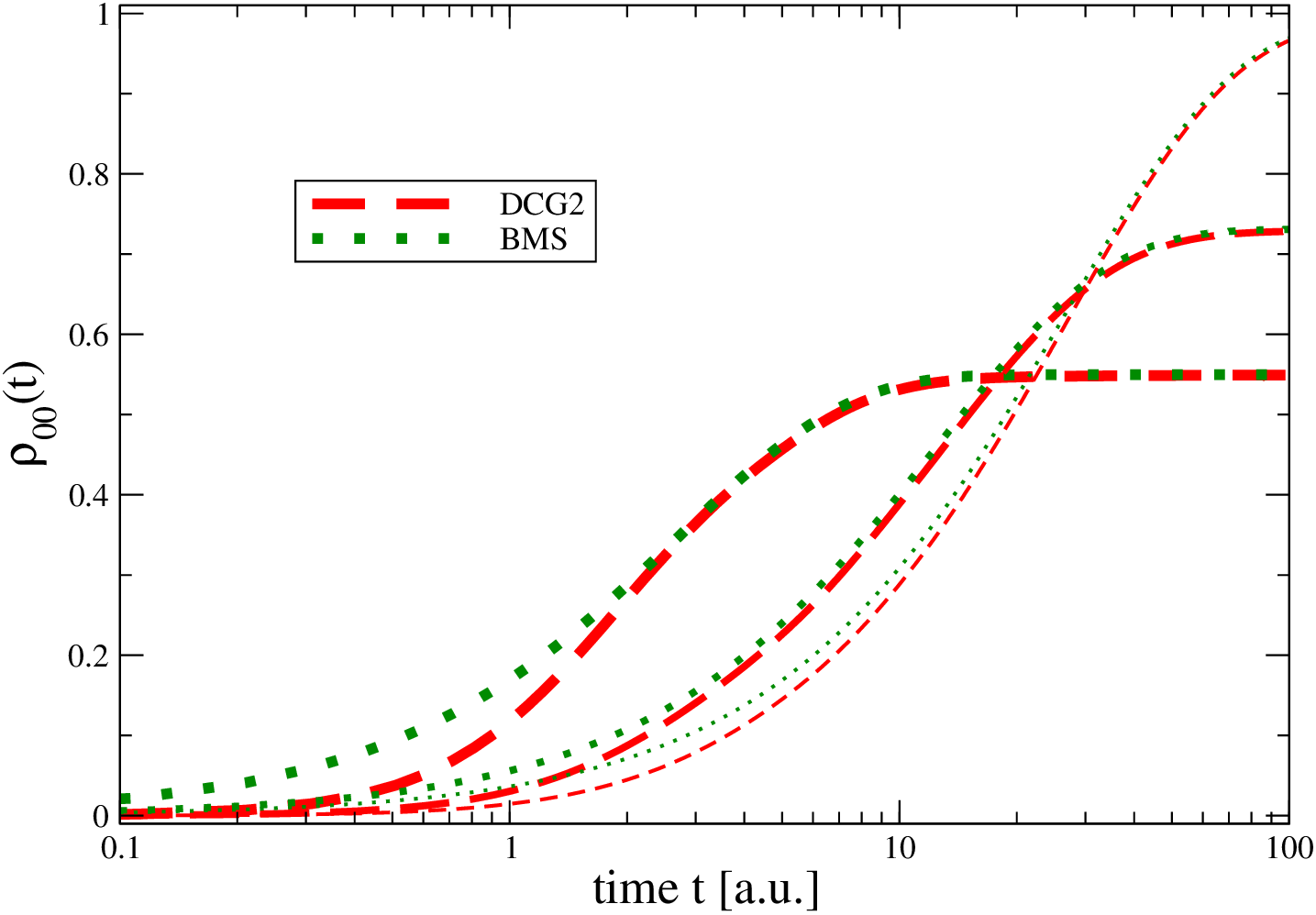}\\
\includegraphics[height=6cm,clip=true]{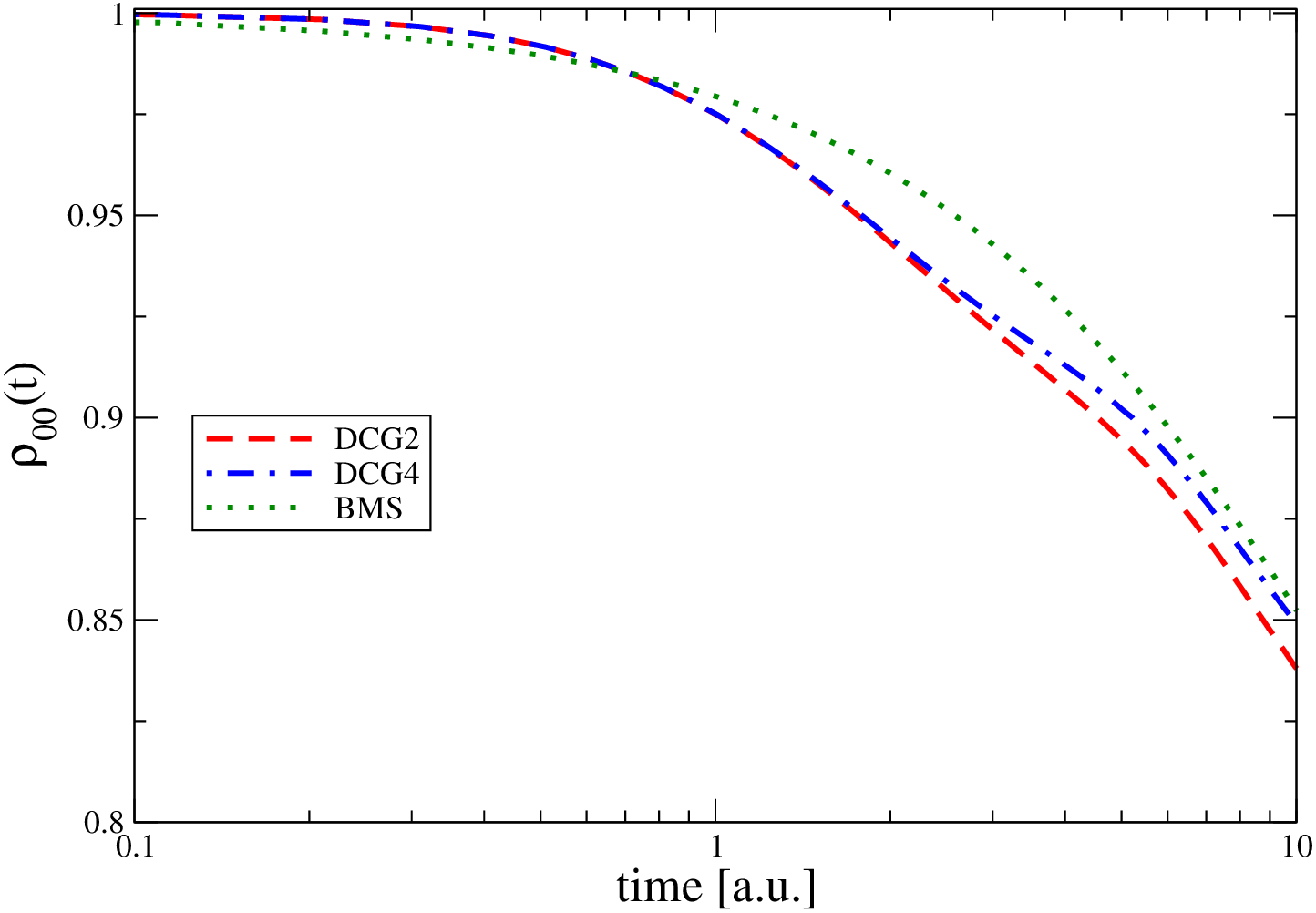}
\end{tabular}
\caption{\label{Fcomp_spinboson}[Color Online]
Comparison of DCG2 (dashed red), DCG4 (dot-dashed blue), and BMS (dotted green) approximations to the dissipative spin-boson model.
In figure (a) we consider different temperatures $\beta=0.2 \ed$ (bold), $\beta=1.0 \ed$ (medium), and
$\beta=5.0\ed$ (thin) in the weak coupling limit $\lambda^2=0.1$ 
and we see that the steady states always coincide whereas for small times DCG and BMS solutions 
differ.
In figure (b) we only show the short time-dynamics for $\lambda^2=0.1$ and $\beta=1.0 \ed$.
The other parameters have been chosen as  $\ed=1$, $\omega_{\rm c}=1$, $G_0=1$, $S=1$.
}
\end{figure}

A general exact solution is unfortunately not available for this case.
However, it is interesting to note that when considering the 
{\em Markov limit} $\beta = 0$, $\omega_{\rm c}\to\infty$, and $S=0$ 
(where the Markov approximation becomes exact)
the correlation function (\ref{Ecfunc2sp}) becomes a $\delta$-function and we see that in this limit, the fourth order
term is cancelled by the squared second order counter term.
For comparison, we plot Born-Markov solution, DCG2, and DCG4 solutions in figure \ref{Fcomp_spinboson}

For the dissipative spin-boson model and Ohmic dissipation ($S=1$) one obtains an exponential decay of the expectation 
value $\expval{\sigma^x}(t)$ in the long time limit \cite{leggett1987a} 
(note the rotation $\sigma^z\to\sigma^x$ and $\sigma^x\to-\sigma^z$).
This corresponds to a decay of the off-diagonal matrix elements and is of course also reproduced by the DCG approach.


\subsection{Fano-Anderson Model}\label{SStransport}

We consider the Fano-Anderson model \cite{fano1961a,anderson1961a}: two leads that are connected by a 
single quantum dot, through which electrons may tunnel from one lead to the other.
The Hamiltonian is given by
\bea\label{Efanoanderson}
H &=& \HS + \HB + \HI\,,\nn
\HS &=& \ed d^\dagger d\,,\;\;
\HB = \sum_{ka} \omega_{ka} c_{ka}^\dagger c_{ka}\,,\nn
\HI &=& \lambda \sum_{ka} \left[t_{ka} d c_{ka}^\dagger - t_{ka}^* d^\dagger c_{ka}\right]
\eea
with fermionic operators creating an alectron with momentum $k$ in 
the left or right lead $a \in\{\rm L, R\}$ for $c_{ka}^\dagger$ or in the
quantum dot $d^\dagger$.
Due to the fermionic anticommutation relations, the model (\ref{Efanoanderson}) 
can be solved exactly for example with Greens functions \cite{haug2008,stefanucci2004a}.
We provide a simplified derivation based on the equation of motion method in appendix \ref{Afanoanderson}.

In contrast to our assumptions in section \ref{Sdcgint}, the operators $d$ and $c_{ka}$ do not act on separate 
Hilbert spaces, which is evident from their anticommutation relations 
$\left\{d, c_{ka}^\dagger\right\}=0$.
This becomes even more explicit via the decomposition
\bea\label{Etransformation}
d &=& \ket{0}\bra{1} \otimes \f{1}\,,\nn
c_{ka} &=& \left[\ket{0}\bra{0} - \ket{1}\bra{1}\right] \otimes \bar c_{ka}\,,
\eea
where $\ket{0}$ and $\ket{1}$ denote the empty and filled dot states, respectively, and the
fermionic operators $\bar c_{ka}$ act only on the (distinct) Fock space of the remaining sites in the leads with
$\left\{ \bar c_{ka}, \bar c_{k'a'}^\dagger \right\}=\delta_{kk'}\delta_{aa'}$.
Naturally, the above decomposition obeys the original anticommutation relations
such as $\left\{d, c_{ka}\right\}=0$ and we conclude for the operators
in the interaction Hamiltonian
\mbox{$d c_{ka}^\dagger = - \ket{0}\bra{1} \otimes \bar c_{ka}^\dagger$}, and
\mbox{$d^\dagger c_{ka} = + \ket{1}\bra{0} \otimes \bar c_{ka}$},
such that now the new operators commute by construction.
Similiar decompositions are also possible systems containing more than one site.
We identify
\bea
A_1 &=& -\ket{0}\bra{1}\,,\qquad A_2 = -\ket{1}\bra{0}\,,\nn
B_1 &=& \sum_{ka} t_{ka} \bar c_{ka}^\dagger\,,\qquad B_2 =  \sum_{ka} t_{ka}^* \bar c_{ka}\,.
\eea
We assume that there are no correlations between left and right leads.
Here we will just consider the infinite bias limit (although this is not crucial, it enables for an analytic calculation of 
all integrals): Taking the chemical potentials to plus or minus infinity for the left and right leads, respectively
($\mu_{\rm L}\to+\infty$ and $\mu_{\rm R}\to-\infty$), we obtain for the fermionic occupation number
\bea
\expval{\bar c_{k\rm L}^\dagger \bar c_{k\rm L}}&=&\frac{1}{e^{\beta(\omega_k-\mu_{\rm L})}+1}\to 1\,,\nn
\expval{\bar c_{k\rm R}^\dagger \bar c_{k\rm R}}&=&\frac{1}{e^{\beta(\omega_k-\mu_{\rm R})}+1}\to 0\,.
\eea
We observe that the coupling operators are non-hermitian in this case.
The correlation functions relevant for the evolution of the diagonal matrix elements can be calculated by introducing continuum 
tunneling rates via 
\mbox{$\Gamma_a(\omega) \equiv 2 \pi \sum_k \abs{t_{ka}}^2 \delta(\omega - \omega_{ka})$}
\bea
C_{\rm L}(t_1-t_2) \equiv C_{12}(t_1, t_2) &=& \frac{1}{2\pi} \int\limits_{-\infty}^{+\infty} \GL(\omega) e^{+i\omega(t_1-t_2)} d\omega\,,\nn
C_{\rm R}(t_2-t_1) \equiv C_{21}(t_1, t_2) &=& \frac{1}{2\pi} \int\limits_{-\infty}^{+\infty} \GR(\omega) e^{-i\omega(t_1-t_2)} d\omega\,,\nn
C_{1212}(t_1, t_2, t_3, t_4) 
&=& C_{\rm L}(t_1-t_2) C_{\rm L}(t_3-t_4)\nn
&& + C_{\rm L}(t_1-t_4) C_{\rm R}(t_3-t_2)\,,\nn
C_{2121}(t_1, t_2, t_3, t_4) 
&=& C_{\rm R}(t_2-t_1) C_{\rm R}(t_4-t_3)\nn
&&+ C_{\rm R}(t_4-t_1) C_{\rm L}(t_2-t_3)\,.
\eea

\begin{widetext}
From Eqn. (\ref{Epertall}), we can derive the second order approximation
\bea
\f{{\cal T}_2^\tau}\RSI &=& \int\limits_0^\tau dt_1 dt_2 \Big[
-C_{21}(t_1,t_2) \Theta(t_2-t_1) \RSI \f{A_1^\dagger}(t_1) \f{A_2^\dagger}(t_2)
-C_{12}(t_1,t_2) \Theta(t_2-t_1) \RSI \f{A_2^\dagger}(t_1) \f{A_1^\dagger}(t_2)\nn
&&
+C_{21}(t_1,t_2) \f{A_1}(t_2) \RSI \f{A_1^\dagger}(t_1)
+C_{12}(t_1,t_2) \f{A_2}(t_2) \RSI \f{A_2^\dagger}(t_1)\nn
&&
-C_{12}(t_1,t_2) \Theta(t_1-t_2) \f{A_1}(t_1) \f{A_2}(t_2) \RSI
-C_{21}(t_1,t_2) \Theta(t_1-t_2) \f{A_2}(t_1) \f{A_1}(t_2) \RSI \Big]\nn
&=& \int\limits_{-\infty}^{+\infty} \frac{d\omega}{2\pi} \GR(\omega) \int\limits_0^\tau dt_1 dt_2
e^{-i(\omega-\ed)(t_1-t_2)}
\left[
-\RSI \ket{1}\bra{1} \Theta(t_2-t_1) + \ket{0}\bra{1}\RSI\ket{1}\bra{0} - \ket{1}\bra{1}\RSI\Theta(t_1-t_2)\right]\nn
&&+\int\limits_{-\infty}^{+\infty} \frac{d\omega}{2\pi} \GL(\omega) \int\limits_0^\tau dt_1 dt_2
e^{+i(\omega-\ed)(t_1-t_2)}
\left[
-\RSI \ket{0}\bra{0} \Theta(t_2-t_1) + \ket{1}\bra{0}\RSI\ket{0}\bra{1} - \ket{0}\bra{0}\RSI\Theta(t_1-t_2)\right]\,,
\eea
from which we can infer the matrix elements of the second-order 
Liouvillian {\em via} $\f{{\cal L}_2^\tau}\RSI = \frac{1}{\tau} \f{{\cal T}_2^\tau}\RSI$.
The Born-Markov-secular approximation is obtained by the Liouvillian ${\cal L}_2^\infty$ with the help of Eqn (\ref{Edeltafunc}).
\end{widetext}

When we parametrize the tunneling rates by Lorentzians \cite{elattari2000a}
\bea\label{Elorentzian}
\Gamma_{\rm R}(\omega) = \frac{\Gamma_{\rm R}^0 \delta_{\rm R}^2}{\left(\omega - \epsilon_{\rm R}\right)^2+\delta_{\rm R}^2}\,,\;\;
\Gamma_{\rm L}(\omega) = \frac{\Gamma_{\rm L}^0 \delta_{\rm L}^2}{\left(\omega - \epsilon_{\rm L}\right)^2+\delta_{\rm L}^2}\,,\qquad
\eea
we obtain an analytic expression for the bath correlation functions in terms of a single decaying exponential
\bea\label{Eexpdecay}
C_{\rm R}(t) = \frac{\Gamma_{\rm R}^0 \delta_{\rm R}}{2} e^{-\abs{t}\delta_{\rm R}+i \epsilon_{\rm R} t}\,,\;\;
C_{\rm L}(t) = \frac{\Gamma_{\rm L}^0 \delta_{\rm L}}{2} e^{-\abs{t}\delta_{\rm L}+i \epsilon_{\rm L} t}\,,\qquad
\eea
such that we can analytically calculate the matrix elements of ${\cal T}_2^\tau$ that govern the evolution of the diagonals
\bea
\left(\begin{array}{cc}
m_{11}(\tau) & m_{14}(\tau)\\
m_{41}(\tau) & m_{44}(\tau)
\end{array}
\right)
&\equiv&
\int\limits_0^\tau dt_1 dt_2 e^{-i \ed (t_1-t_2)}\times\nn
&&\left(\begin{array}{cc}
-C_{\rm L}(t_1-t_2) & +C_{\rm R}(t_1-t_2)\\
+C_{\rm L}(t_1-t_2) & -C_{\rm R}(t_1-t_2)
\end{array}
\right)\,.
\eea
Then, we obtain from $\bra{0} \f{{\cal T}_2^\tau}\RSI \ket{0}$ for the evolution of the diagonal matrix element
\bea
\f{\dot{\rho}_{00}^\tau} &=& 
\lambda^2 \left[\frac{m_{11}(\tau)}{\tau} \f{\rho_{00}^\tau}(t) + \frac{m_{14}(\tau)}{\tau} \f{\rho_{11}^\tau}(t)\right]\,,
\eea
where we can exploit trace conservation 
\mbox{$\f{\rho_{11}^\tau}(t)=1-\f{\rho_{00}^\tau}(t)$} (this feature is trivially fulfilled
by the coarse-graining approach). 
Afterwards, the above equation can explicitly be solved for 
\bea
\f{\rho_{00}^\tau}(t) &=& \f{\rho_{00}}(0) \exp\left\{\lambda^2\left[m_{11}(\tau)-m_{14}(\tau)\right]\frac{t}{\tau}\right]\nn
&&+\frac{1-\exp\left\{\lambda^2\left[m_{11}(\tau)-m_{14}(\tau)\right]\frac{t}{\tau}\right\}}{1-\frac{m_{11}(\tau)}{m_{14}(\tau)}}
\eea

\begin{widetext}
For the fourth-order contribution we obtain (extensively using  
$C_{\bar{1}\bar{2}\bar{1}\bar{2}}(t_1,t_2,t_3,t_4)=C_{2121}(t_1,t_2,t_3,t_4)$, 
$C_{1112}(t_1,t_2,t_3,t_4)=0$ etc.)
\bea
\f{{\cal T}_4^\tau}\RSI &=& \int\limits_0^\tau dt_1 dt_2 dt_3 dt_4 \Big[
+C_{2121}(t_1, t_2, t_3, t_4)\Theta(t_4 - t_3) \Theta(t_3 - t_2)\Theta(t_2 - t_1)
\RSI \f{A_1^\dagger}(t_1) \f{A_2^\dagger}(t_2) \f{A_1^\dagger}(t_3) \f{A_2^\dagger}(t_4)\nn
&&
+C_{1212}(t_1, t_2, t_3, t_4)\Theta(t_4 - t_3) \Theta(t_3 - t_2)\Theta(t_2 - t_1)
\RSI \f{A_2^\dagger}(t_1) \f{A_1^\dagger}(t_2) \f{A_2^\dagger}(t_3) \f{A_1^\dagger}(t_4)\nn
&&
-C_{2121}(t_1, t_2, t_3, t_4)\Theta(t_3 - t_2) \Theta(t_2 - t_1)
\f{A_1}(t_4) \RSI \f{A_1^\dagger}(t_1) \f{A_2^\dagger}(t_2) \f{A_1^\dagger}(t_3)\nn
&&
-C_{1212}(t_1, t_2, t_3, t_4)\Theta(t_3 - t_2) \Theta(t_2 - t_1)
\f{A_2}(t_4) \RSI \f{A_2^\dagger}(t_1) \f{A_1^\dagger}(t_2) \f{A_2^\dagger}(t_3)\nn
&&
+C_{2112}(t_1, t_2, t_3, t_4)\Theta(t_3 - t_4) \Theta(t_2 - t_1)
\f{A_1}(t_3) \f{A_2}(t_4) \RSI \f{A_1^\dagger}(t_1) \f{A_2^\dagger}(t_2)\nn
&&
+C_{2121}(t_1, t_2, t_3, t_4)\Theta(t_3 - t_4) \Theta(t_2 - t_1)
\f{A_2}(t_3) \f{A_1}(t_4) \RSI \f{A_1^\dagger}(t_1) \f{A_2^\dagger}(t_2)\nn
&&
+C_{1212}(t_1, t_2, t_3, t_4)\Theta(t_3 - t_4) \Theta(t_2 - t_1)
\f{A_1}(t_3) \f{A_2}(t_4) \RSI \f{A_2^\dagger}(t_1) \f{A_1^\dagger}(t_2)\nn
&&
+C_{1221}(t_1, t_2, t_3, t_4)\Theta(t_3 - t_4) \Theta(t_2 - t_1)
\f{A_2}(t_3) \f{A_1}(t_4) \RSI \f{A_2^\dagger}(t_1) \f{A_1^\dagger}(t_2)\nn
&&
-C_{2121}(t_1, t_2, t_3, t_4)\Theta(t_2 - t_3) \Theta(t_3 - t_4)
\f{A_1}(t_2) \f{A_2}(t_3) \f{A_1}(t_4) \RSI \f{A_1^\dagger}(t_1)\nn
&&
-C_{1212}(t_1, t_2, t_3, t_4)\Theta(t_2 - t_3) \Theta(t_3 - t_4)
\f{A_2}(t_2) \f{A_1}(t_3) \f{A_2}(t_4) \RSI \f{A_2^\dagger}(t_1)\nn
&&
+C_{1212}(t_1, t_2, t_3, t_4)\Theta(t_1 - t_2) \Theta(t_2 - t_3)\Theta(t_3 - t_4)
\f{A_1}(t_1) \f{A_2}(t_2) \f{A_1}(t_3) \f{A_2}(t_4) \RSI\nn
&&
+C_{2121}(t_1, t_2, t_3, t_4)\Theta(t_1 - t_2) \Theta(t_2 - t_3)\Theta(t_3 - t_4)
\f{A_2}(t_1) \f{A_1}(t_2) \f{A_2}(t_3) \f{A_1}(t_4) \RSI
\Big]\,.
\eea
The relevant part in above equation for the evolution of the diagonals evaluates by using Eqn. (\ref{Eheavi00}) to
\bea
\left(\f{{\cal T}_4^\tau} \RSI\right)_{11} &\equiv & p_{11}(\tau)  \f{\rho_{00}^\tau}(t) + p_{14}(\tau) \f{\rho_{11}^\tau}(t)\,,\nn
p_{11}(\tau) &=&
+\int\limits_0^\tau dt_1 dt_2 dt_3 dt_4 e^{-i \ed(+t_1-t_2+t_3-t_4)}\times\nn
&&\left[C_{\rm L}(t_1-t_2) C_{\rm L}(t_3-t_4) + C_{\rm L}(t_1-t_4) C_{\rm R}(t_3-t_2)\right]
\left[\Theta(t_3-t_2) \Theta(t_2-t_1) + \Theta(t_2-t_3) \Theta(t_3-t_4)\right]\qquad\nn
p_{14}(\tau) &=&
-\int\limits_0^\tau dt_1 dt_2 dt_3 dt_4 e^{+i \ed(t_1-t_2+t_3-t_4)}\times\nn
&&\left[C_{\rm R}(t_2-t_1) C_{\rm R}(t_4-t_3) + C_{\rm R}(t_4-t_1) C_{\rm L}(t_2-t_3)\right]
\left[\Theta(t_3-t_2) \Theta(t_2-t_1) + \Theta(t_2-t_3) \Theta(t_3-t_4)\right]\,,
\eea
where again all integrals can be solved analytically yielding even lengthier expressions than before.
\end{widetext}
Together with
\bea
\bra{0} \f{{\cal T}_2^\tau} \f{{\cal T}_2^\tau} \RSI \ket{0} &=&
\left[m_{11}(\tau) m_{11}(\tau) + m_{14}(\tau) m_{41}(\tau)\right] \f{\RS^{00}}\nn
&+&\left[m_{11}(\tau) m_{14}(\tau) + m_{14}(\tau) m_{44}(\tau)\right] \f{\RS^{11}}\nonumber
\eea
we obtain from
\bea
\lambda^2 \f{{\cal L}^\tau_2} + \lambda^4 \f{{\cal L}^\tau_4} &=& 
\frac{\lambda^2}{\tau} \f{{\cal T}_2^\tau} 
+ \frac{\lambda^4}{\tau}\left(\f{{\cal T}_4^\tau}
- \frac{\tau^2}{2} \f{{\cal L}^\tau_2} \f{{\cal L}^\tau_2}\right)\nn
&=& 
\frac{1}{\tau}\left(\lambda^2 \f{{\cal T}_2^\tau} + \lambda^4 \f{{\cal T}_4^\tau} - \frac{\lambda^4}{2} \f{{\cal T}_2^\tau} \f{{\cal T}_2^\tau} \right)
\eea
the following differential equation for the diagonals
\bea
\f{\dot{\rho}_{00}^\tau} &=& \frac{1}{\tau} \Big\{
\lambda^2 m_{11}(\tau)- \frac{\lambda^4}{2}\Big[m_{11}(\tau) m_{11}(\tau)\nn
&&+ m_{14}(\tau) m_{41}(\tau)\Big]+\lambda^4 p_{11}(\tau) \Big\}\f{\rho_{00}^\tau}(t)\nn
&&+\frac{1}{\tau} \Big\{\lambda^2 m_{14}(\tau) -\frac{\lambda^4}{2} \Big[m_{11}(\tau) m_{14}(\tau)\nn
&&+ m_{14}(\tau) m_{44}(\tau)\Big]+\lambda^4 p_{14}(\tau) \Big\} \f{\rho_{11}^\tau}(t)\,.
\eea
The above equation can be explicitly solved for $\f{\rho_{00}^\tau}(\tau)$ by imposing trace conservation as in subsection 
\ref{SSspinboson}.
It can be shown analytically that in the flatband limit $\delta_{\rm R}\to\infty, \delta_{\rm L}\to\infty$ (where for infinite bias 
the Markovian approximation becomes exact), the fourth order correction in the above equation is cancelled by the counterterm from the 
second order for all graining times $\tau$.
Moreover, one can also show analytically that the apparent divergence for large graining 
times $p_{11}\propto \tau^2$ and $p_{14}\propto \tau^2$ 
is precisely cancelled by the fourth order counter terms arising from the second order.

\begin{figure}
\includegraphics[height=6cm,clip=true]{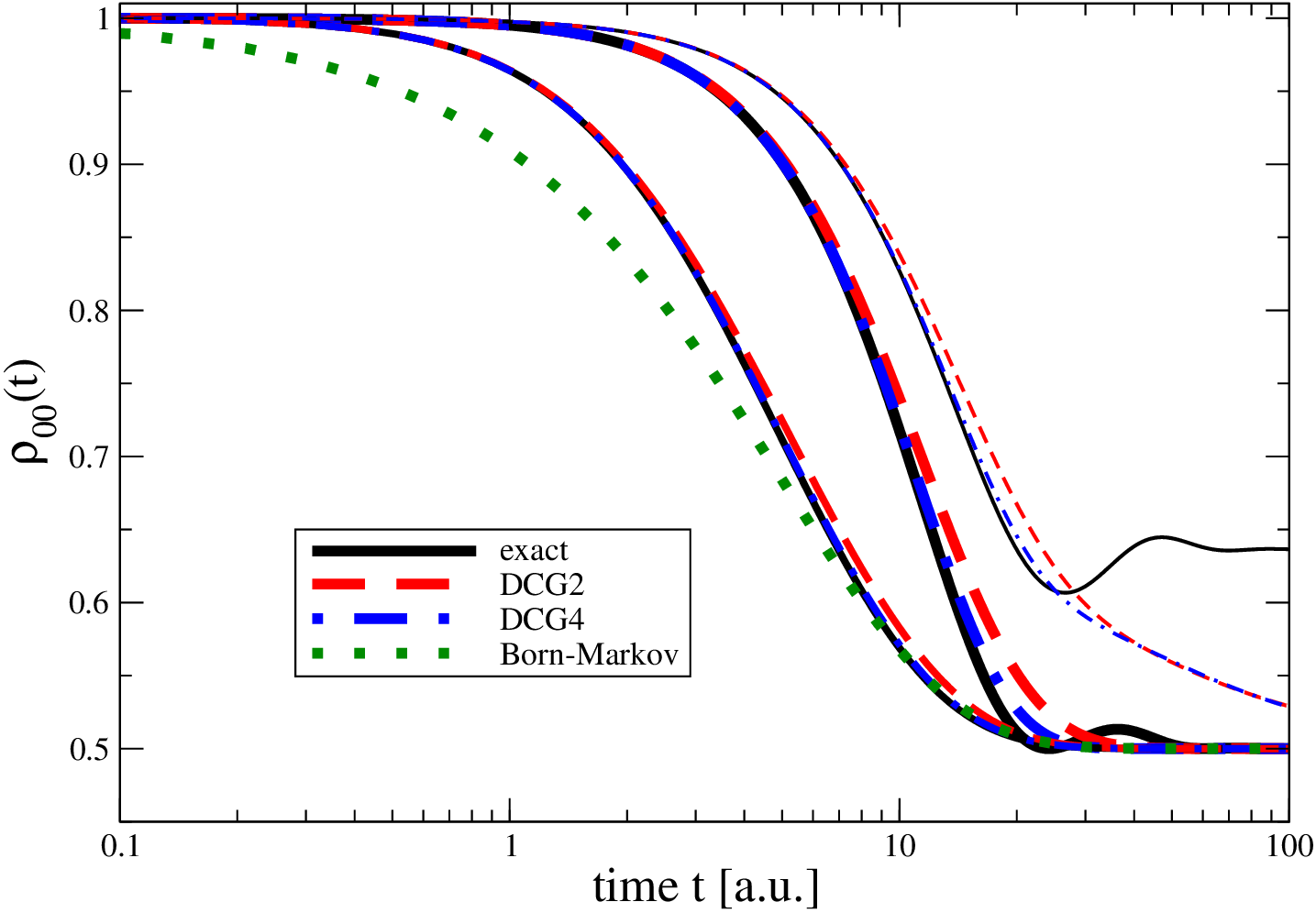}
\caption{\label{Fcomp_philipp0}[Color Online]
Comparison of exact (solid black), DCG2 (dashed red), DCG4 (dot-dashed blue), and Born-Markov 
(dotted green, same for all parameters) solutions for the 
Fano-Anderson model.
The bold lines ($\dl=\dr=0.1$) show the highly non-Markovian regime whereas the medium thickness lines ($\dl=\dr=1.0$)
denote a regime where the Markovian approximation performs comparably well.
The thin lines ($\dr=2\dl=0.1$) demonstrate the failure of the DCG solutions in the large-time limit.
The other parameters have been chosen as $\lambda^2=0.1$, $\GR^0=\GL^0=1$, and $\er=\ed=\el=1$.
}
\end{figure}

\begin{figure}
\begin{tabular}{c}
\includegraphics[height=6cm,clip=true]{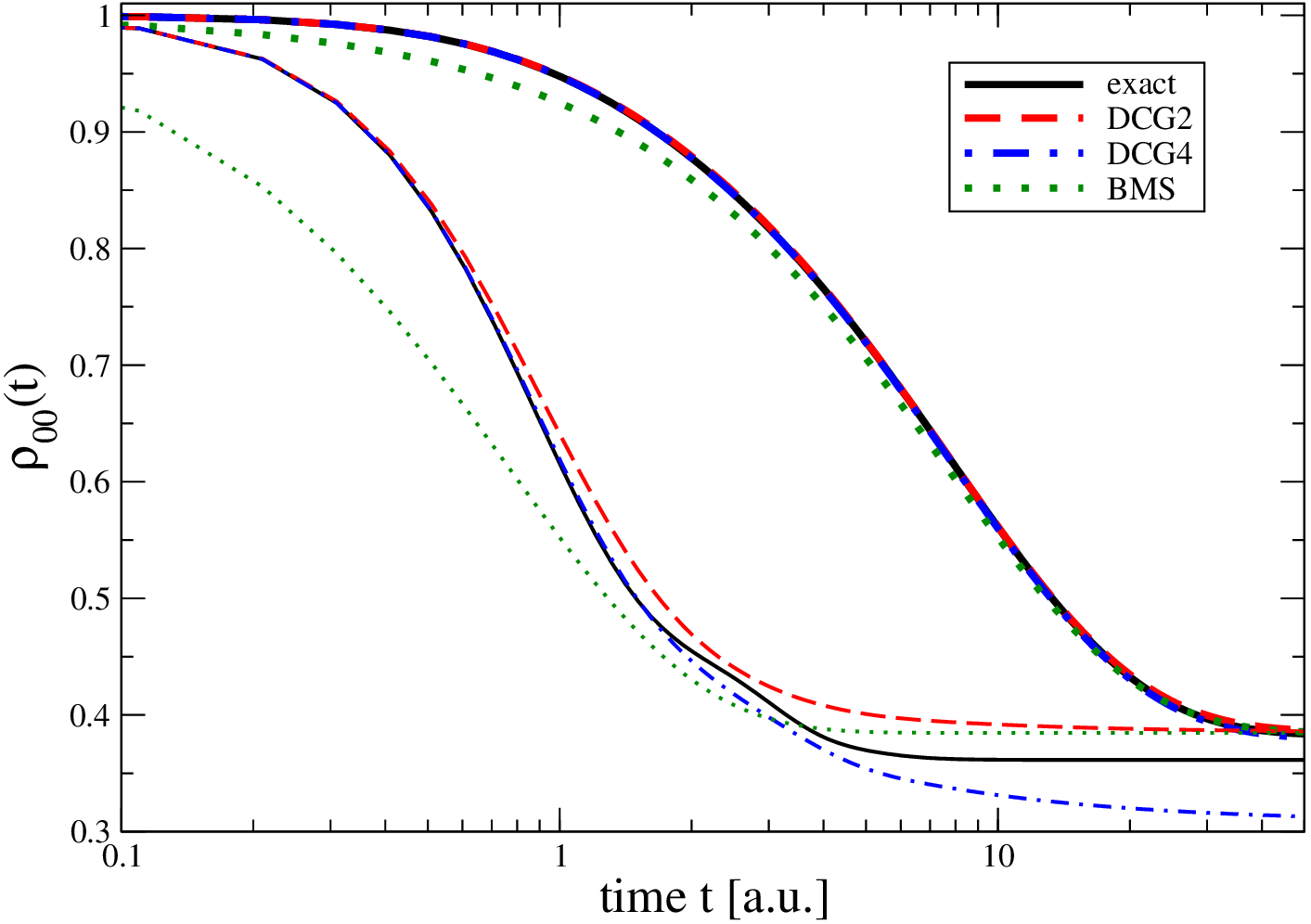}\\
\includegraphics[height=6cm,clip=true]{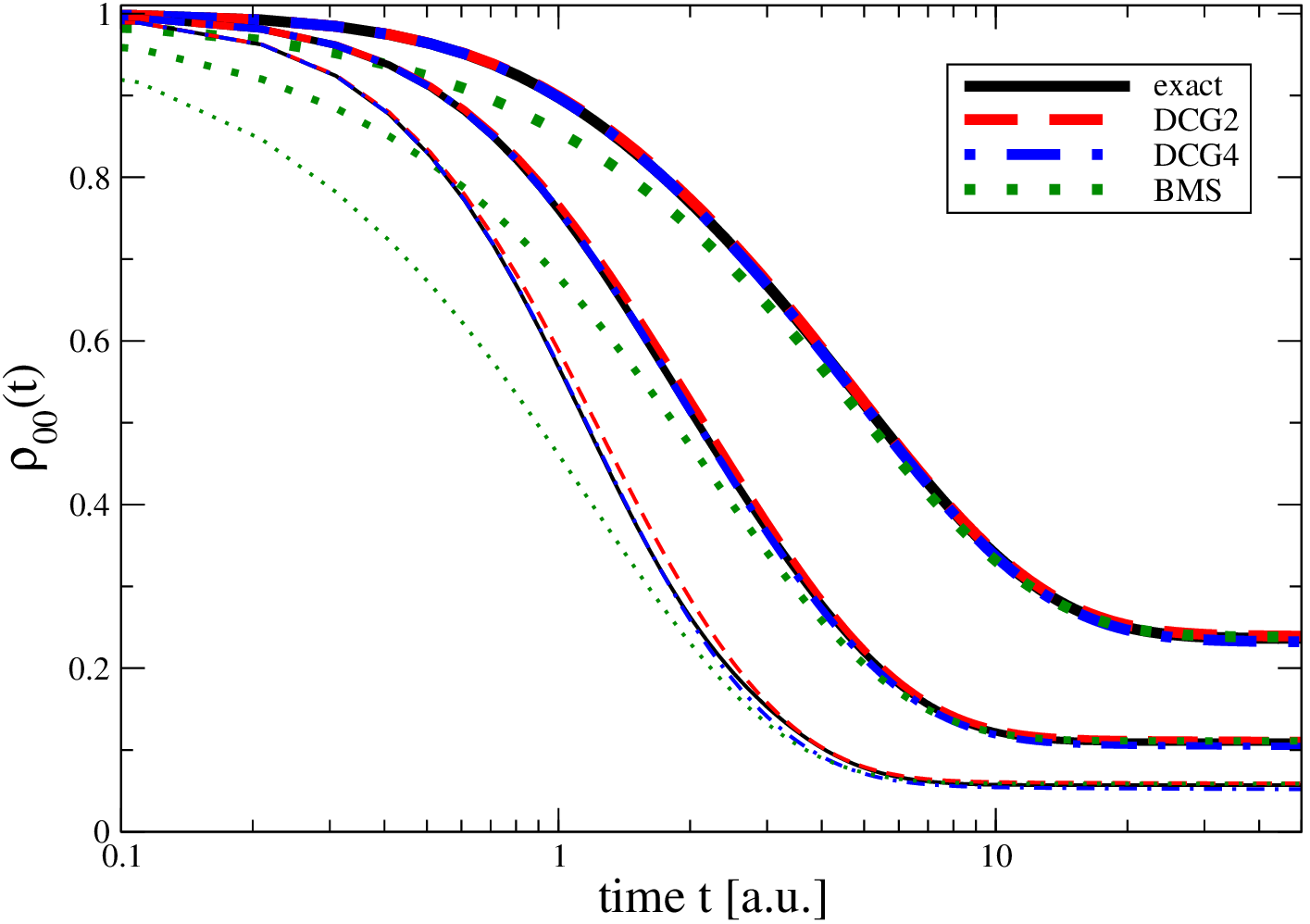}
\end{tabular}
\caption{\label{Fcomp_philipp}[Color Online]
Comparison of exact (solid black), DCG2 (dashed red), DCG4 (dot-dashed blue), and Born-Markov (dotted green) 
solutions for the Fano-Anderson model.
In figure (a), we have considered symmetric maximum tunneling rates $\GL^0=\GR^0=1$ for the
weak coupling limit $\lambda^2=0.1$ (bold lines) and the strong coupling limit $\lambda^2=1.0$ (thin lines).
It is visible that the steady state of BMS and DCG2 solutions does not depend on the coupling strength and in
the strong coupling limit, the steady state of DCG4 might actually be worse than that of DCG2.
In figure (b), we have used $\lambda^2=0.1$, $\GR^0=1.0$ and varied the left maximum tunneling rate as
$\GL^0=2.0$ (bold), $\GL^0=5.0$ (medium), and $\GL^0=10.0$ (thin).
For small times, DCG4 always yields a better result than DCG2 and both DCG solutions are better than
the BMS approximation.
The latter performs particularly bad for small times as expected.
The other parameters have (in both figures) been chosen as:
$\dr=1$, $\dl=2$, $\er=\el=0$, $\ed=1$.
}
\end{figure}

In contrast, one obtains under the Born-Markov (the secular approximation has no effect for 
this particular example) approximation the solution
\bea
\f{\rho_{00}}(\tau) &=& \frac{\GR(\ed)}{\GL(\ed)+\GR(\ed)}\left(1- e^{-\lambda^2 \left[\GL(\ed)+\GR(\ed)\right] \tau}\right)\nn
&&+ \f{\rho_{00}}(0) e^{-\lambda^2 \left[\GL(\ed)+\GR(\ed)\right] \tau}\,,
\eea
which has been derived using Eqn. (\ref{Emarkovlimit}) and identity (\ref{Edeltafunc}).
From the Lorentzian tunneling rates (\ref{Elorentzian}) we see that the Markovian solution is completely 
independent on the width of the tunneling rates $\dr$, $\dl$, see also figure \ref{Fcomp_philipp0}.
Especially for small widths $\dr$, $\dl$ the correlation functions (\ref{Eexpdecay}) decay very slowly and we also
observe a large difference between Born-Markov and exact solution, whereas the DCG solutions perform comparably well.
With the exact solution from appendix \ref{Afanoanderson} we plot the Born-Markov solution, and DCG2 as well as DCG4 solutions
in figure \ref{Fcomp_philipp} for varying coupling strengths as well as for different model symmetries.


\section{Summary}

The dynamical coarse-graining approach has been extended to higher orders.
By performing the derivation in the interaction picture, we have directly demonstrated that
the DCG solution must approximate the exact solution by construction for small times.
This has been confirmed by several examples.
Interestingly, the DCG method even reproduced the complete recurrences of the diagonal density 
matrix elements in case of small reservoirs.
For continuous reservoirs, the short time dynamics of DCG4 has always been superior to the
short-time performance of DCG2. 
This however need not be the case for the large time limit.
However, the performance of DCG2 is always better (short time) or equal (large time) to the
performance of the Born-Markov-secular approximation.

We have shown that DCG2 unconditionally preserves positivity.
Going beyond \cite{schaller2008a}, this also includes reservoirs that are not in equlibrium.
In addition, the presentation in the interaction picture leads to a much simpler form of DCG2,
such that now it appears at least as simple (if not simpler) than the conventional Born-Markov theory.
Positivity is not unconditionally preserved for higher orders of DCG.

Unfortunately, the DCG method is computationally quite demanding in the interesting case of large 
(continuous) reservoirs, as it requires the evaluation of high-dimensional integrals.
As the dimension of some integrals can be reduced by analytical integration, the efficiency of DCG
may therefore strongly depend on the structure of the bath correlation functions.
For systems that are larger than the single spins considered here, it will also prove difficult to 
calculate the exponential of ${\cal L}^\tau t$ for all $t$ and $\tau$ with moderate computational effort.
If however the result is only of interest at a specific time (as is the case for example in adiabatic quantum
computation) one only has to exponentiate a single matrix or -- even simpler -- evolve the density matrix 
according to a single Liouvillian.
The fact that DCG2 unconditionally preserves positivity also for time-dependent system Hamiltonians
renders the method a good candidate for analyzing the corrections of decoherence to adiabatic quantum
computation \cite{farhi2001} without the necessity
of reverting to conventional Born-Markov-secular theory \cite{childs2001a}.


\section{Acknowledgments}

G. S. gratefully acknowledges discussions with J. Eisert, C. Emary, G. Kiesslich, 
R. Sch\"utzhold, and M. Vogl.


\begin{appendix}


\section{Approximate Steady State of the Spin-Boson model}\label{Aspinboson}

Starting from the Hamiltonian (\ref{Ehamspinboson}) we obtain with the abbreviations (we
use bold symbols for operators in the Heisenberg picture)
\bea
\f{s_k} = h_k \f{b_k} + h_k^* \f{b_k}^\dagger\,,\qquad \f{a_k} = i(h_k \f{b_k} - h_k^* \f{b_k}^\dagger)
\eea
a set of equations for the time-evolutions of $\expval{\f{\sigma^x}}$, $\expval{\f{\sigma^y}}$, $\expval{\f{\sigma^z}}$, 
$\expval{\f{\sigma^x s_k}}$, $\expval{\f{\sigma^x a_k}}$, $\expval{\f{\sigma^y s_k}}$, $\expval{\f{\sigma^y a_k}}$, 
$\expval{\f{\sigma^z s_k}}$, $\expval{\f{\sigma^z a_k}}$ in the Heisenberg picture, which is unfortunately not closed.
However, we can achieve closure by assuming factorization of the expectation values and stationarity
of the reservoir
\bea
\expval{\f{\sigma^{x/y/z}} \sum_{k'}(\f{s_{k'} s_k + s_k s_{k'}})} &=& 2 \expval{\f{\sigma^{x/y/z}}}\expval{\f{s_k}^2}\,,\nn
\expval{\f{s_k}^2} &=& \abs{h_k}^2\left[1+2n(\omega_k)\right]\,,\nn
\expval{\f{\sigma^{x/y/z}} \sum_{k'}(\f{s_{k'} a_k + a_k s_{k'}})} &=& 0\,,\nn
\expval{\sum_{k'}(\f{s_{k'} a_k - a_k s_{k'}})} &=& -2i \abs{h_k}^2\,,
\eea
which is consistent with the Born approximation.
We can then analyze the steady state of the resulting equations and obtain the same results as
discussed before, i.e., for pure dephasing ($A=\sigma^z$) we obtain
$\expval{\f{\sigma^x}_\infty}=\expval{\f{\sigma^y}_\infty}=0$ and $\expval{\f{\sigma^z}_\infty}=\expval{\sigma^z_0}$
as discussed in subsection \ref{SSpuredephasing} 
and similarly for the dissipative case ($A=\sigma^x$) we obtain 
$\expval{\f{\sigma^x}_\infty}=\expval{\f{\sigma^y}_\infty}=0$ and 
$\expval{\f{\sigma^z}_\infty}=\frac{1-e^{-\beta\ed}}{1+e^{-\beta\ed}}$, 
which is consistent with Eqn. (\ref{Ess_sb}) in subsection \ref{SSdissipation}.


\section{Exact solution of the Fano-Anderson model for Lorentzian tunneling rates}\label{Afanoanderson}

From the Hamiltonian (\ref{Efanoanderson}) we can calculate the time evolution of the fermionic operators in 
the Heisenberg picture (we use bold operator symbols to denote the Heisenberg picture and exploit the
anticommutation relations throughout)
\bea
\dot{\f{d}} &=& -i \ed \f{d} + i \lambda \sum_k \left[ t_{k\rm L}^* \f{c_{k\rm L}} + t_{k\rm R}^* \f{c_{k\rm R}}\right]\,,\nn
\dot{\f{c_{k\rm L}}} &=& -i \omega_{k\rm L} \f{c_{k\rm L}} + i \lambda t_{k\rm L} \f{d}\,,\nn
\dot{\f{c_{k\rm R}}} &=& -i \omega_{k\rm R} \f{c_{k\rm R}} + i \lambda t_{k\rm R} \f{d}\,,
\eea
which already forms a closed set of equations.
These equations can be Laplace-transformed (where $\f{d}(t) \to \tilde{d}(z)$ and 
$\dot{\f{d}}(t) \to -d + z \tilde{d}(z)$ and similarly for the other operators).
\begin{widetext}
In Laplace space, we can eliminate $\tilde{c}_{k\rm L}(z)$ and $\tilde{c}_{k\rm R}(z)$ to solve the resulting equations for 
\bea
\tilde{d}(z) &=& \frac{d + i \lambda \sum_k \left(
\frac{t_{k\rm L}^* c_{k\rm L}}{z + i \omega_{k\rm L}} + \frac{t_{k\rm R}^* c_{k\rm R}}{z + i\omega_{k\rm R}}\right)}
{z + i\ed + \lambda^2 \sum_k \left(\frac{\abs{t_{k\rm L}}^2}{z+i\omega_{k\rm L}} + \frac{\abs{t_{k\rm R}}^2}{z+i\omega_{k\rm R}}\right)}
= \frac{d + i \lambda \sum_k \left(
\frac{t_{k\rm L}^* c_{k\rm L}}{z + i \omega_{k\rm L}} + \frac{t_{k\rm R}^* c_{k\rm R}}{z + i\omega_{k\rm R}}\right)}
{z + i\ed + \frac{\lambda^2}{2\pi} \int\limits_{-\infty}^{+\infty} \frac{\GL(\omega) + \GR(\omega)}{z+i\omega}d\omega}\nn
&=& \frac{(z+\dl + i\el)(z+\dr+i\er)\left[
d + i \lambda \sum_k \left(
\frac{t_{k\rm L}^* c_{k\rm L}}{z + i \omega_{k\rm L}} + \frac{t_{k\rm R}^* c_{k\rm R}}{z + i\omega_{k\rm R}}\right)\right]}
{(z + i\ed)(z+\dl+i\el)(z+\dr+i\er) + \frac{\lambda^2}{2} \left[\GL^0 \dl (z+\dr+i\er) 
+ \GR^0 \dr (z+\dl+i\el)\right]}\,,
\eea
where in the last line we have already assumed Lorentzian tunneling rates (\ref{Elorentzian}).
The inverse Laplace transform (Bromwick integral \cite{arfken2005}) can be performed by the theorem of residues 
$\f{d}(t)=\left.\sum_i {\rm Res} \tilde{d}(z) e^{+z t}\right|_{z=z_i}$, 
where $z_i$ denote the poles of $\tilde{d}(z)$.
Denoting the roots of
\bea
(z + i\ed)(z+\dl+i\el)(z+\dr+i\er) + \frac{\lambda^2}{2} \left[\GL^0 \dl (z+\dr+i\er) 
+ \GR^0 \dr (z+\dl+i\el)\right] = (z-z_1)(z-z_2)(z-z_3)
\eea
by $z_1$, $z_2$, and $z_3$, respectively, we can easily calculate the residues (In case 
of degenerate roots, one may either use residue formulae for higher-order poles or simply
analytic continuation of the solution for first order poles).
For $z_1\neq z_2, z_1 \neq z_3$, and $z_2 \neq z_3$ we obtain the solution
\bea
\f{d}(t) &=& \Big[
+\frac{(z_1+\dl+i\el)(z_1+\dr+i\er) e^{z_1 t}}{(z_1-z_2)(z_1-z_3)}
+\frac{(z_2+\dl+i\el)(z_2+\dr+i\er) e^{z_2 t}}{(z_2-z_1)(z_2-z_3)}\nn
&&+\frac{(z_3+\dl+i\el)(z_3+\dr+i\er) e^{z_3 t}}{(z_3-z_1)(z_3-z_2)}
\Big] d\nn
&&+i\lambda\sum_k\Big[
+\frac{(z_1+\dl+i\el)(z_1+\dr+i\er) e^{z_1 t}}{(z_1-z_2)(z_1-z_3)(z_1+i\omega_{k\rm L})}
+\frac{(z_2+\dl+i\el)(z_2+\dr+i\er) e^{z_2 t}}{(z_2-z_1)(z_2-z_3)(z_2+i\omega_{k\rm L})}\nn
&&+\frac{(z_3+\dl+i\el)(z_3+\dr+i\er) e^{z_3 t}}{(z_3-z_1)(z_3-z_2)(z_3+i\omega_{k\rm L})}
+\frac{(-i\omega_{k\rm L}+\dl+i\el)(-i\omega_{k\rm L}+\dr+i\er) e^{-i\omega_{k\rm L} t}}
{(-i\omega_{k\rm L}-z_1)(-i\omega_{k\rm L}-z_2)(-i\omega_{k\rm L}-z_3)}\Big]t_{k\rm L}^* c_{k\rm L}\nn
&&+i\lambda\sum_k\Big[
+\frac{(z_1+\dl+i\el)(z_1+\dr+i\er) e^{z_1 t}}{(z_1-z_2)(z_1-z_3)(z_1+i\omega_{k\rm R})}
+\frac{(z_2+\dl+i\el)(z_2+\dr+i\er) e^{z_2 t}}{(z_2-z_1)(z_2-z_3)(z_2+i\omega_{k\rm R})}\nn
&&+\frac{(z_3+\dl+i\el)(z_3+\dr+i\er) e^{z_3 t}}{(z_3-z_1)(z_3-z_2)(z_3+i\omega_{k\rm R})}
+\frac{(-i\omega_{k\rm R}+\dl+i\el)(-i\omega_{k\rm R}+\dr+i\er) e^{-i\omega_{k\rm R} t}}
{(-i\omega_{k\rm R}-z_1)(-i\omega_{k\rm R}-z_2)(-i\omega_{k\rm R}-z_3)}\Big]t_{k\rm R}^* c_{k\rm R}\,.
\eea
With taking the initial conditions as $\expval{c_{k'\rm L}^\dagger c_{k\rm L}}=\delta_{kk'} f_{\rm L}(\omega_{k\rm L})$ and
$\expval{c_{k'\rm R}^\dagger c_{k\rm R}}=\delta_{kk'} f_{\rm R}(\omega_{k\rm R})$ and $\expval{c_{k'\rm R}^\dagger c_{k\rm L}}=0$
we obtain for $n(t)=\expval{\f{d^\dagger}(t)\f{d}(t)}$ the expression
\bea\label{Esolexactfa}
n(t) &=& 
\left|
\frac{(z_1+\dl+i\el)(z_1+\dr+i\er) e^{z_1 t}}{(z_1-z_2)(z_1-z_3)}
+\frac{(z_2+\dl+i\el)(z_2+\dr+i\er) e^{z_2 t}}{(z_2-z_1)(z_2-z_3)}\right.\nn
&&\left.+\frac{(z_3+\dl+i\el)(z_3+\dr+i\er) e^{z_3 t}}{(z_3-z_1)(z_3-z_2)}\right|^2 n_0\nn
&&+\frac{\lambda^2}{2\pi}\int\limits_{-\infty}^{+\infty} 
\left[\GL(\omega) f_{\rm L}(\omega) + \GR(\omega) f_{\rm R}(\omega)\right]\times\nn
&&
\left|
\frac{(z_1+\dl+i\el)(z_1+\dr+i\er) e^{z_1 t}}{(z_1-z_2)(z_1-z_3)(z_1+i\omega)}
+\frac{(z_2+\dl+i\el)(z_2+\dr+i\er) e^{z_2 t}}{(z_2-z_1)(z_2-z_3)(z_2+i\omega)}\right.\nn
&&\left.+\frac{(z_3+\dl+i\el)(z_3+\dr+i\er) e^{z_3 t}}{(z_3-z_1)(z_3-z_2)(z_3+i\omega)}
+\frac{(-i\omega+\dl+i\el)(-i\omega+\dr+i\er) e^{-i\omega t}}
{(-i\omega-z_1)(-i\omega-z_2)(-i\omega-z_3)}\right|^2 d\omega\,.
\eea
\end{widetext}
In the large time-limit, this considerably simplifies (with using that $\Re z_i < 0$).
Conventionally, $\lambda^2$ is absorbed in $\GL(\omega)$ and $\GR(\omega)$ and by setting
$\lambda\to 1$ we explicitly recover the well-known steady-state results in the literature 
(compare e.g., Eqns (12.27) with using Lorentzian tunneling rates of form (\ref{Elorentzian}), 
Eqns. (12.30), and (12.31) in ref. \cite{haug2008} with 
$n_\infty=\frac{-i}{2\pi} \int\limits_{-\infty}^{+\infty} G^<(\omega) d\omega$).


\end{appendix}


\end{document}